# ALKEMIE Agent: an autonomous platform for computational materials design

Hongfu Huang[1,2], Yuzhe Li[2], Ao Xu[2], Bo Liu[2], Changrui Wang[2], Kan Tang[2], Ning Yang[2], Shengxian Liu[2], Hanyu Liu[2], Pengpeng Zhang[2], Linggang Zhu[2*], Fengkai Liu[3], Yichen Lu[2], Tong Zhao[2], Naihua Miao[2], Jian Zhou[2*] and Zhimei Sun[2*]

[1]School of Integrated Circuit Science and Engineering, Beihang University, Beijing, 100191, China.

[2]School of Materials Science and Engineering, Beihang University, Beijing, 100191, China.

[3]School of Computer Science and Engineering, Beihang University, Beijing, 100191, China.

[*]Corresponding author:

lgzhu7@buaa.edu.cn; jzhou@buaa.edu.cn; zmsun@buaa.edu.cn

**Abstract**

Despite the powerful multi-scale modeling methods and high-throughput infrastructures established in the materials community, real material computation workflows remain fragmented and heavily manual, requiring researchers to constantly bridge software tools, data analysis, and intermediate decisions. This growing gap between methodological capability and practical execution highlights the need for a new kind of autonomous computational framework, one that can coordinate tools, knowledge, and workflows in a more unified and adaptive way. Here, we introduce ALKEMIE Agent, an agentic platform in which retrieval-augmented generation, a materials-computation knowledge base, registered skills, database-supported provenance, AI-assisted structure modeling, bounded task execution, tool-calling iteration, and error-diagnostic assistance are integrated within a traceable control loop. The capabilities of ALKEMIE Agent are demonstrated through applications including materials recommendation, structure modeling, phonon calculations, machine-learned interatomic potential training, LAMMPS simulations, Ab Initio Monte Carlo (AIMC) sampling, and active-learning-based materials screening. Finally, we outline the future directions and challenges for the development of agentic platforms for computational materials design.

## 1. Introduction

Computer simulation has long been regarded as a third paradigm of scientific discovery, complementing experiment and theory, and has increasingly converged with data-driven approaches. It enables the investigation of complex scientific problems that are difficult, costly, or even infeasible to address through experiments and theory. Consequently, in the field of materials science, computational materials design (CMD) has become an essential route for decoding the structure-property relationship, accelerating the discovery and deployment of novel materials. Over multiple length and time scales, a wide spectrum of methods, including density functional theory (DFT), molecular dynamics (MD), machine-learned interatomic potentials (MLIPs), high-throughput calculations, phase-field modeling, and finite-element-based continuum simulations, have enabled predictive modeling of materials behavior from the electronic to the component scale with increasing accuracy and efficiency[1–11]. Over the past decades, this progress has been further strengthened by transformative frameworks such as Integrated Computational Materials Engineering (ICME) and the Materials Genome Initiative/Engineering (MGI/MGE), which have promoted the development of multi-scale simulation strategies, materials databases, workflow automation engines, and high-throughput infrastructures. Notably, high-throughput infrastructures such as pymatgen[12], ALKEMIE[13], AiiDA[14], AFLOW[15], OQMD[16], Atomate[17], and FireWorks[18] have established robust ecosystems for structure processing, automated calculations, data management, and computational workflow orchestration. Among these infrastructures, ALKEMIE, which stands for Artificial Learning and Knowledge Enhanced Materials Informatics Engineering, established an integrated computational platform for data generation, data management, and data mining by combining high-throughput calculation, automated computational structure-model construction, machine-learning analysis, machine-learned interatomic potential generation, and graphical workflow and dataflow management.

Despite these advances, practical computational materials design is still difficult to organize as an adaptive and traceable task chain. Existing databases, high-throughput engines, and calculation-management platforms are effective for standardized data generation and repeated execution; however, many real-world research tasks still require context-dependent decisions across heterogeneous software, files, parameters, and intermediate results. For example, a candidate material identified from a database or screening model often requires transformation into a simulation-ready object through supercell construction, surface or interface modeling, defect introduction, substitutional editing, format conversion, input generation, and consistency checking before a solver can be executed. Similarly, expert intervention is frequently required during method selection, job submission, execution monitoring, error diagnosis, result parsing, and database update. As a result, the bottleneck in CMD has shifted from the high-throughput execution of predefined calculations to the supervised coordination of task definition, domain knowledge, computational tools, intermediate artifacts, and reproducible records.

The rapid development of large language models (LLMs) and tool-augmented AI agents provides a route for addressing part of the above-mentioned challenges by combining natural-language task interpretation with retrieval, external tool use, reusable skills, memory, and traceable intermediate actions[19–32]. Recently, materials scientists have begun exploring LLM-based and agent-based frameworks for computational materials research. Existing proposed computational materials agents can be broadly categorized by functionality: retrieval-grounded systems, such as HoneyComb[26] and LLaMP[33], emphasize materials knowledge access, database retrieval, and tool-enhanced reasoning; solver-centered systems, such as VASPilot[25] and MDAgent[31], focus on specific simulation environments and input or script generation; multi-task frameworks, such as MatSciAgent[24] and Matty[27], integrate materials databases, structure generation, molecular simulation, continuum simulation, or workflow construction; and design-oriented or multi-agent systems, including

MatExpert[28], S1-MatAgent[29], TopoMAS[23], and AtomAgents[34], incorporate decomposition strategies, recommendation, physics-aware simulation, multimodal analysis, or domain-specific validation in selected materials problems. Although these systems demonstrate the feasibility of agent-assisted materials computation, their capabilities remain largely fragmented, often restricted to knowledge retrieval, single-code execution, selected materials domains, or target-specific design loops. A more general and extensible platform is needed to connect materials recommendation, AI-assisted structure modeling and editing, software-specific execution, error-diagnosis assistance, task records, self-iteration and database-supported provenance within a human-supervised autonomy boundary.

To address this gap, ALKEMIE Agent is introduced as a human-supervised agentic platform for computational materials design. Building on the graphical interface, high-throughput workflow management, and multi-scale computational capabilities of ALKEMIE[13], the platform connects natural-language task specification, materials recommendation, AI-assisted structure modeling and editing, retrieval-augmented generation (RAG), a materials-computation knowledge base, a skill library, context and memory management, database-supported provenance, active-learning-based iteration, and bounded execution for selected VASP, LAMMPS, Gradient-optimized Neuroevolution Potential (GNEP)[9], and Ab Initio Monte Carlo (AIMC) routes. Autonomy is operationally limited to observable platform actions, including task decomposition, tool selection, input preparation, execution preparation or submission, monitoring, file and log inspection, result parsing, diagnostic assistance, memory update, and human-approved next-step suggestion. The architecture, core modules, and representative applications of the platform are described below.

## 2. Design overview of ALKEMIE Agent

### 2.1 Key capabilities

As discussed above, ALKEMIE Agent is designed to support the transition from fixed automation pipelines to human-supervised agentic coordination. Fig. 1 summarizes the conceptual basis for this transition, including design objectives, layered architecture, and control flow.

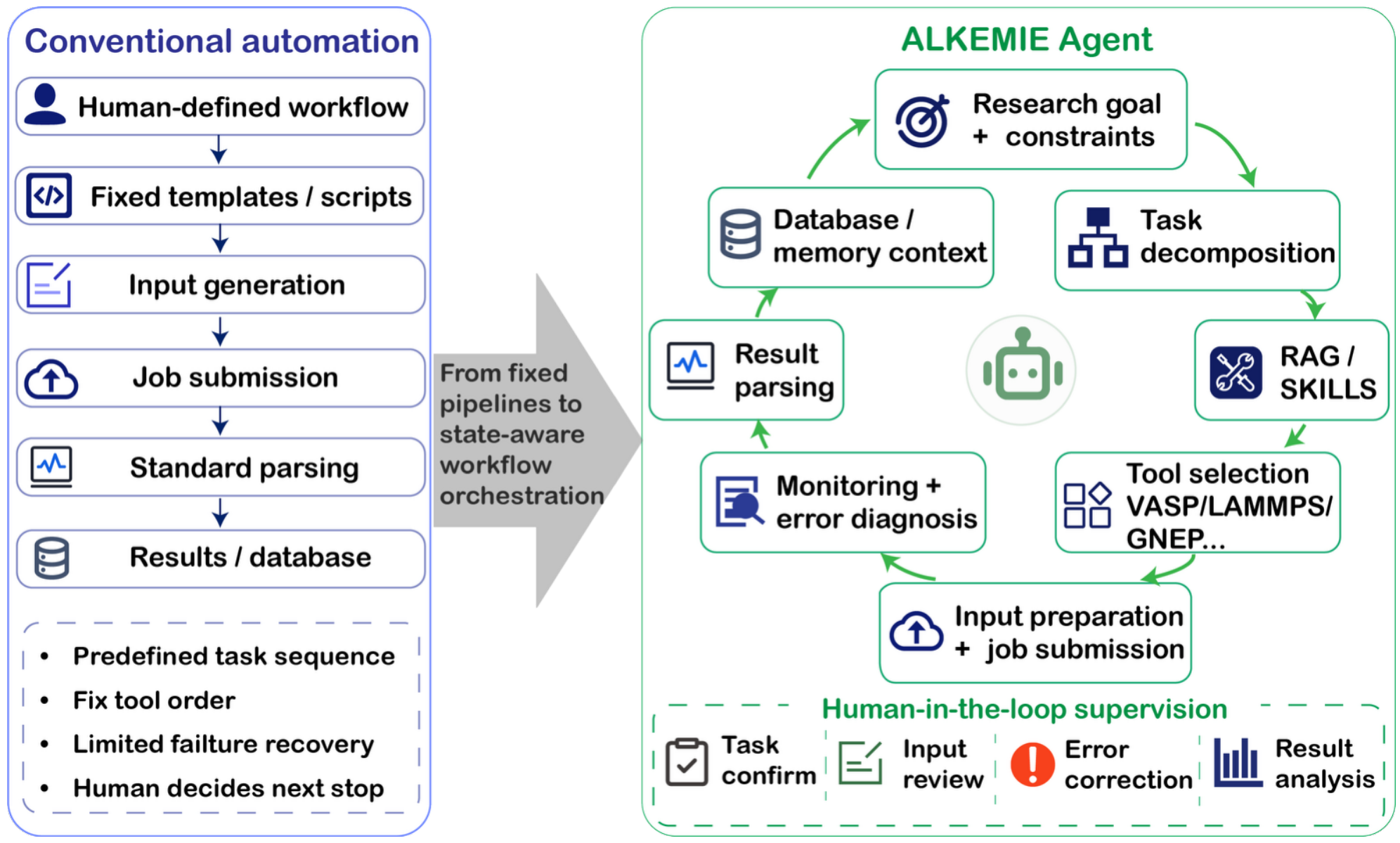

Fig. 1. Transition from fixed automation pipelines to agentic task coordination in computational materials design. ALKEMIE Agent reorganizes structure preparation, input generation, execution, analysis, and data recording into a human-supervised agentic control loop supported by RAG, registered skills, bounded task execution, execution traces, and database-supported provenance.

At the platform level, as shown in Fig. 2, ALKEMIE Agent is organized around four functional groups: user-facing task interaction, domain-grounded reasoning, solver-facing execution, and traceable data/provenance management. The user-facing workspace integrates task specification, agent interaction, structure visualization, task orchestration, and result inspection. Domain-grounded reasoning is supported by the knowledge base, RAG module, and skill library. Solver-facing execution is connected to structure modeling, software interfaces, monitoring, parsing, and diagnosis. Data and provenance management retain files, task states, parsed outputs, artifacts, and execution records. Together, these groups support the conversion of research objectives into knowledge-grounded, executable, inspectable, and traceable computational actions.

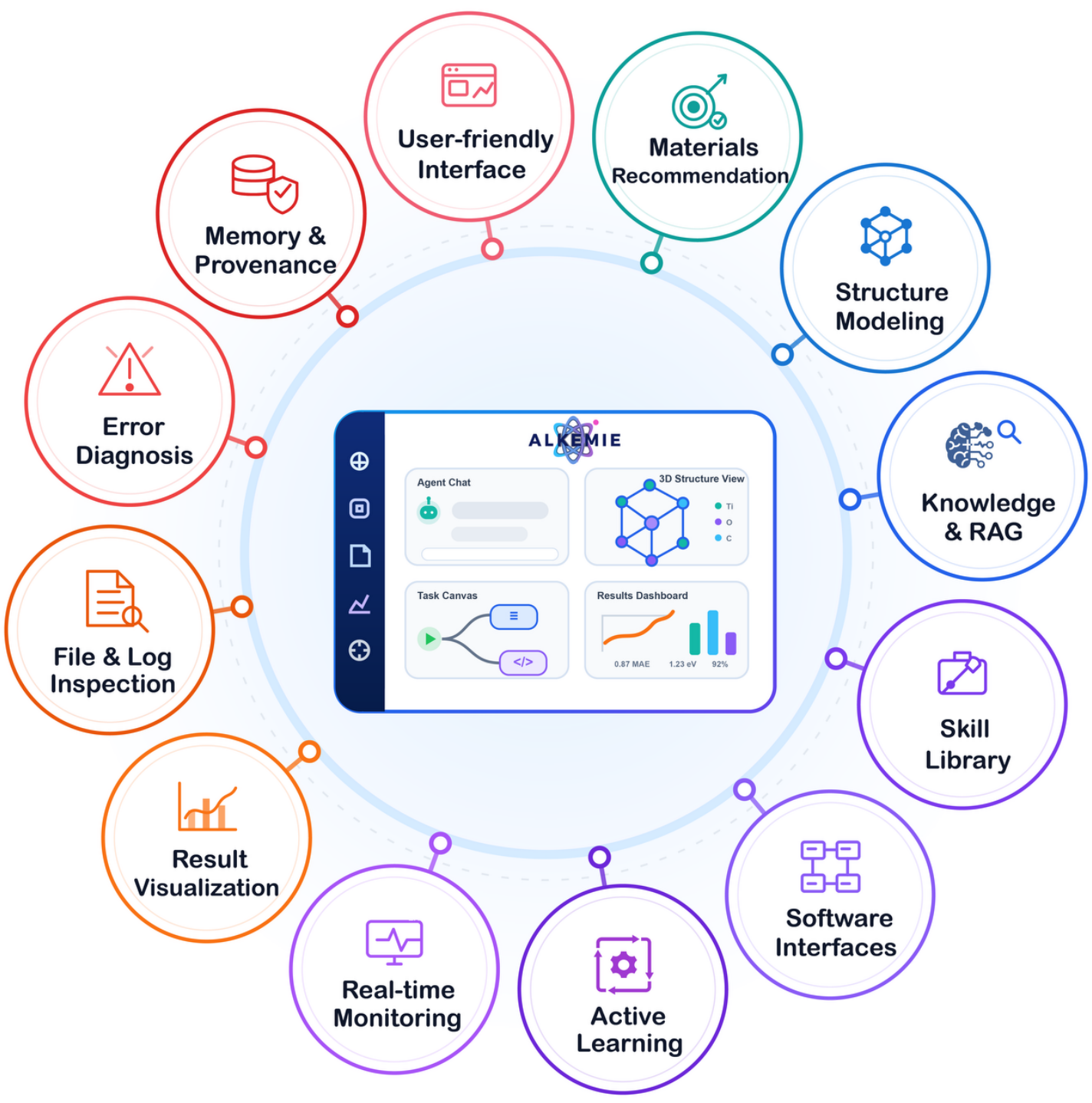

Fig. 2. Core functions of ALKEMIE Agent. The platform connects a user-facing workspace with functional modules for domain-grounded reasoning, materials-object preparation, solver-facing execution, and traceable data/provenance management.

## 2.2 Bounded Autonomy

Autonomy in ALKEMIE Agent is defined within an explicit operational boundary. In ALKEMIE Agent, natural-language task descriptions are interpreted into structured intents, relevant context is assembled, and executable computational routes are proposed, with support for initiating preparation and post-processing actions. Importantly, key calculation assumptions are made explicit and remain visible prior to execution, ensuring transparency and controllability throughout the workflow. This boundary is required because choices of structure model, solver, exchange-correlation functional, interatomic potential, convergence threshold, sampling protocol, and scheduler configuration may alter the physical interpretation of a result. Autonomy in ALKEMIE Agent is therefore defined as supervised coordination of route proposal, input preparation, status inspection, result parsing, and next-step suggestion, rather than unrestricted model-driven scientific decision making.

This autonomy boundary also determines how reproducibility and domain knowledge are implemented. A supervised action is treated as meaningful only when the selected route, generated artifacts, execution state, raw outputs, parsed quantities, analysis artifacts, and provenance records remain available for inspection and reproducibility. For this reason, the agent layer is connected to task states, database records, file management, output parsing, and event traces. Solver-specific settings, materials-specific conventions, and task-specific modeling requirements are supported by retrieval-augmented generation, knowledge-based entries, and skill-library components. The intended scope of ALKEMIE Agent is consequently defined as human-supervised coordination within documented computational routes and traceable execution boundaries, rather than open-ended autonomous discovery.

### 2.3 Architecture

ALKEMIE Agent is organized as a layered architecture that connects human-facing interaction, agentic reasoning, knowledge, skills, task execution, and data/provenance management. These layers are implemented through a client-server-infrastructure architecture, in which browser-based access is connected to backend agent and execution services, while computational jobs are handled by HPC resources. As summarized in Fig. 3, the architecture provides a common environment in which task specification, route coordination, software access, and provenance recording are linked rather than treated as separate operations.

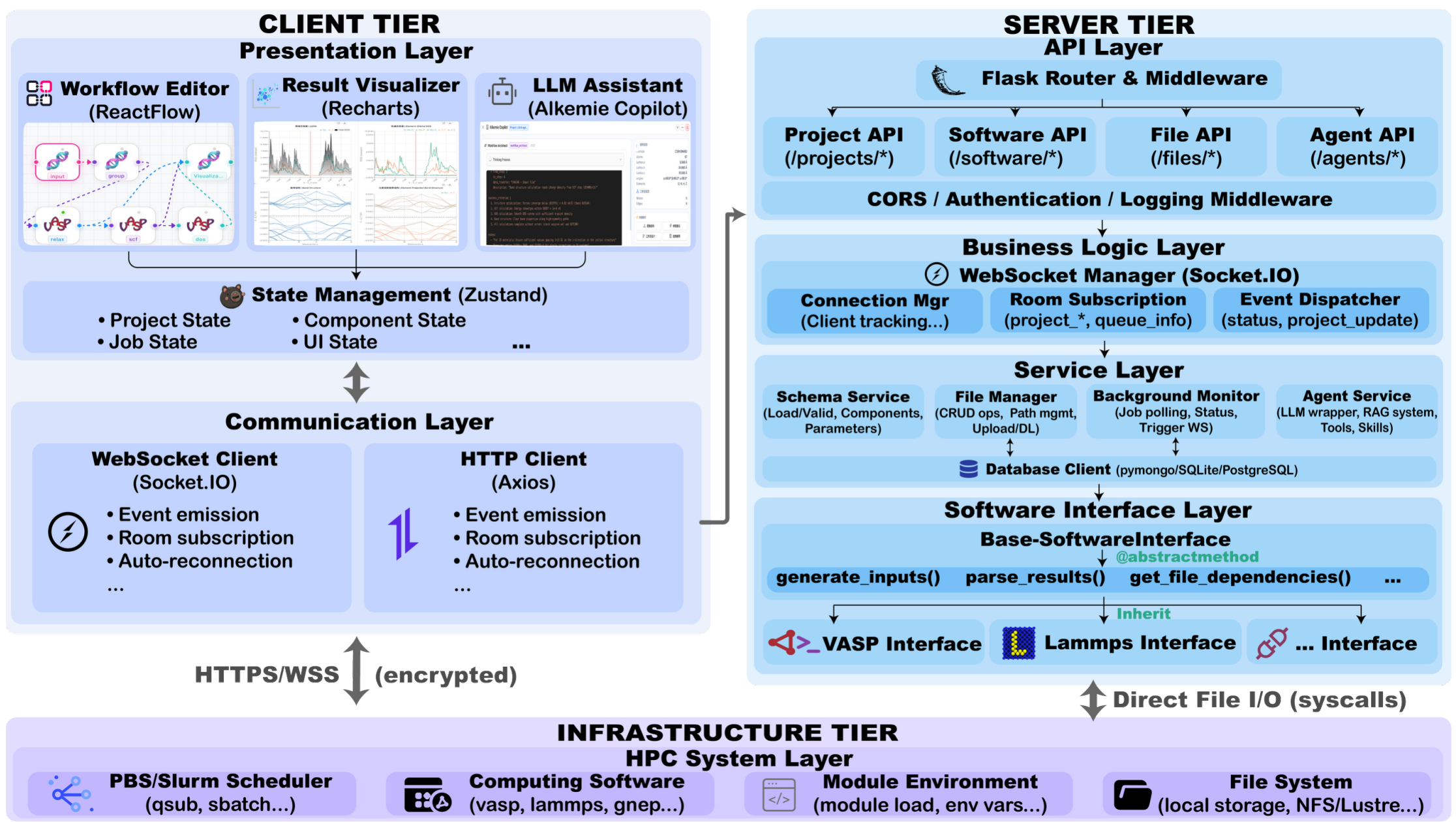

Fig. 3. Client-server-infrastructure architecture of ALKEMIE Agent. The browser-based workspace, backend agent services, execution services, HPC resources, and data/provenance layer are connected to support task specification, route coordination, software execution, and record preservation.

As for the human-agent interaction, natural-language requests, uploaded structures or files, calculation parameters, and user confirmations are associated with the active project state. Intermediate responses, task status, calculation progress, parsed results, and analysis outputs are returned through the same workspace, so that execution-relevant assumptions and approvals remain visible before computational actions are carried out. The collected context is then passed to the agentic reasoning layer, where the request is classified and mapped to an appropriate service route. According to the task state, the request may be directed to task-route construction, parameter assistance, scientific interpretation, component analysis, knowledge retrieval, structure-related operation, or software-specific execution. The runtime components involved in this routing process are shown in Fig. 4.

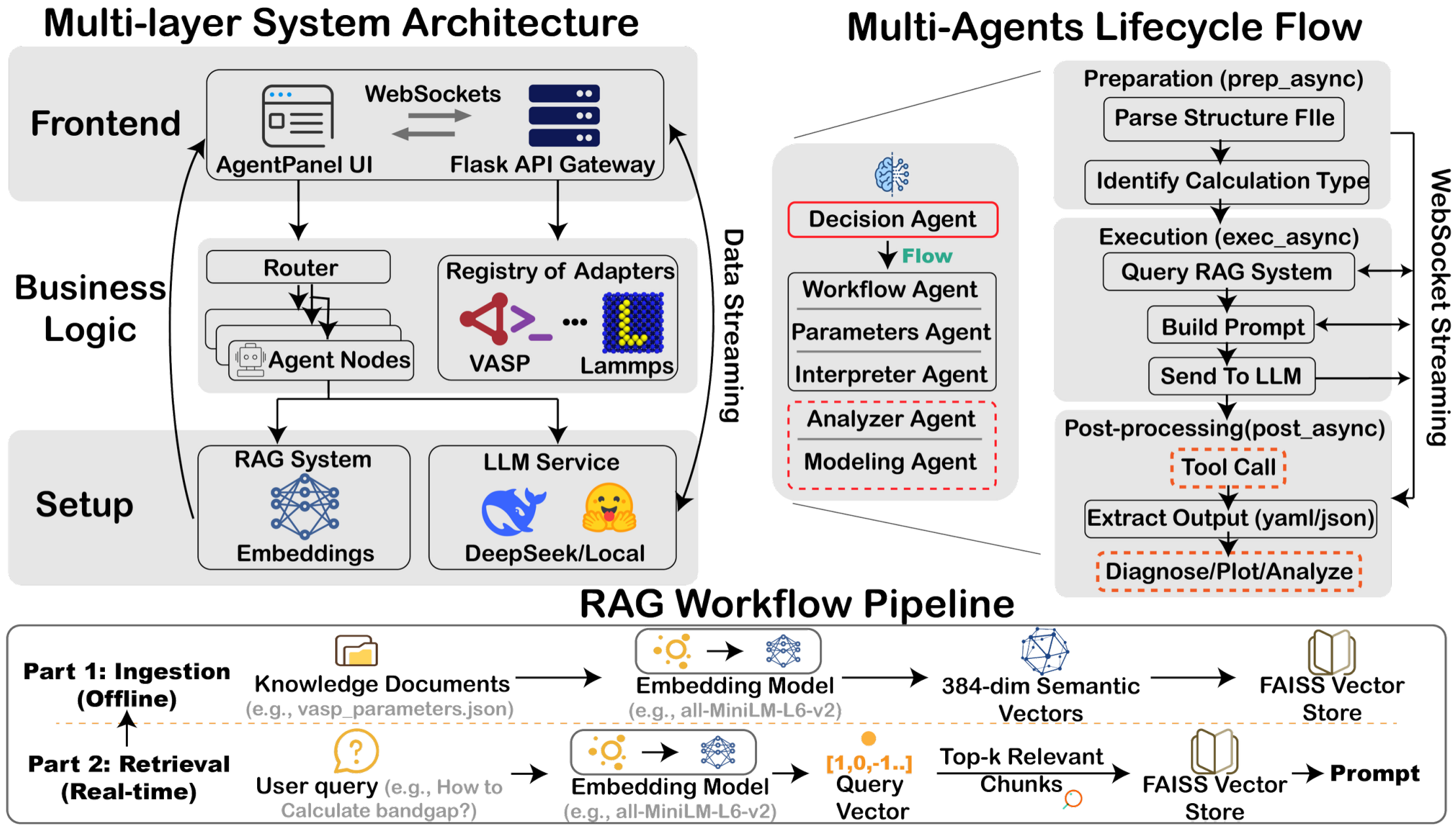

Fig. 4. Agent-runtime components and information flow. The panel shows how task interpretation, retrieval, skill selection, software-specific execution, and result parsing are connected inside the platform.

The knowledge and skill layer provides domain expertise for the reasoning process, whereas the task execution layer connects validated routes to supported computational software and execution resources. The data/provenance layer retains task states, generated files, job metadata, parsed outputs, analysis artifacts, user feedback, and agent traces as reconstructable records. The internal mechanisms of these resources, including knowledge retrieval, memory and provenance, structure modeling, software interfaces, tool-calling iteration, and active learning, are described in Section 3.

## 2.4 Agentic control loop and execution trace

The layered architecture is operationalized as a bounded, human-supervised control loop in which a materials-design request is converted into traceable platform states. As summarized in Fig. 5, a task objective is first interpreted against project context, uploaded files, active structures, task status, and user-provided constraints. Missing information and execution-relevant assumptions are identified during consultation, after which a computational route, solver choice, or analysis path is proposed for review. Following confirmation, the required structures, files, parameters, and prerequisites are checked before the route is passed to the corresponding backend service or software interface. In this arrangement, autonomy is expressed through task-route construction, context assembly, tool selection, preparation, monitoring, parsing, and human-approved next-step suggestion, while calculation-defining choices remain available for inspection before execution.

The execution trace provides the record through which this control loop is made reproducible. Knowledge retrieval, skill selection, structure preparation, input generation, execution preparation, status monitoring, file or log inspection, parser invocation, and analysis are associated with the active project, task node, files, jobs, parser outputs, and artifacts. Intermediate observations may be returned to the active context before later operations are proposed. The event-level mechanism supporting this process is described in Section 3.3.

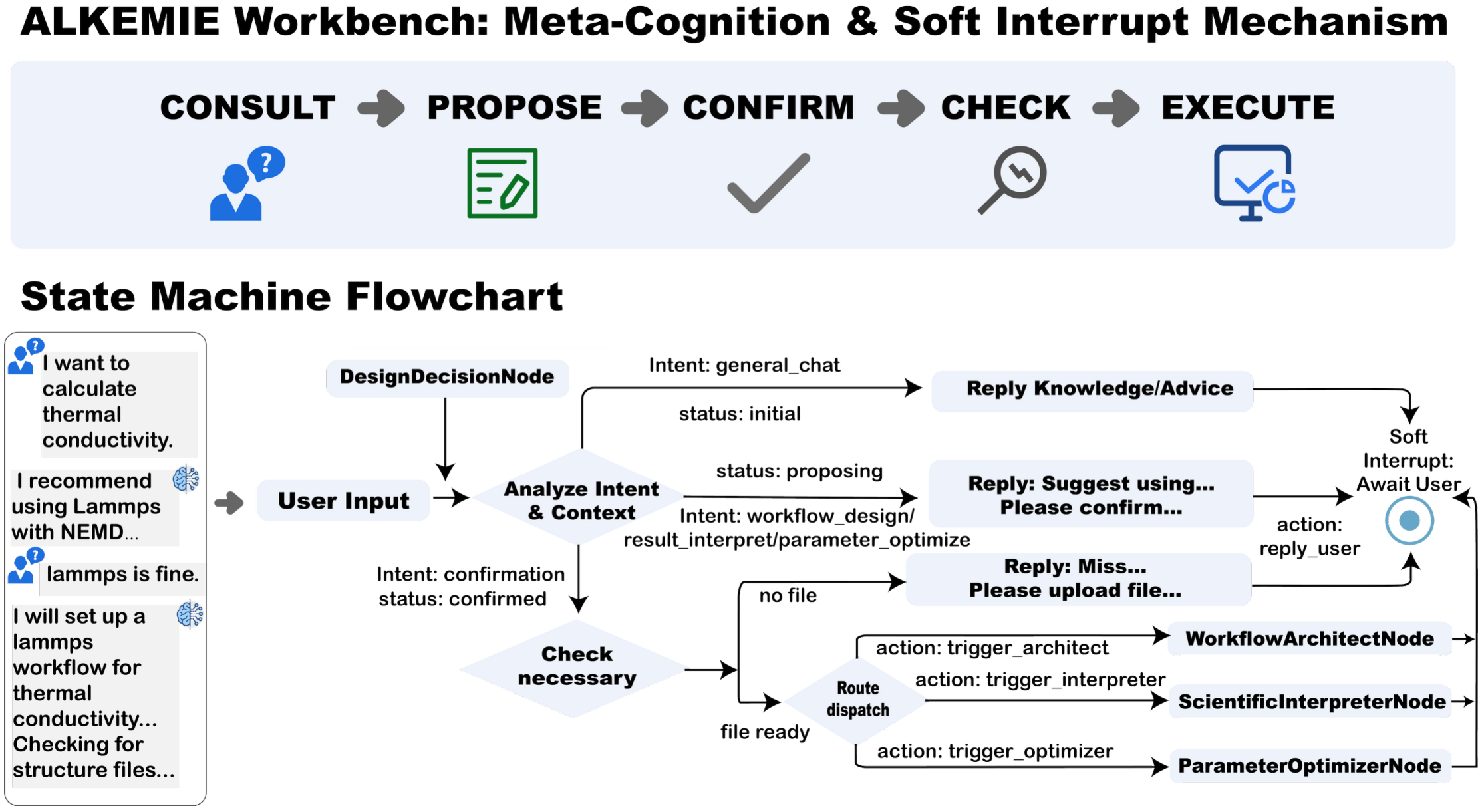

Fig. 5. Human-supervised agentic control loop in ALKEMIE Agent.

Failure handling and result closure are represented through the same trace structure. When abnormal status records, parser outputs, or log patterns are detected, available evidence is assembled as diagnostic context and returned as a reviewable suggestion. Any subsequent correction is recorded only when it is invoked through a permitted interface. After successful execution, raw outputs are parsed into structured quantities, plots, or analysis artifacts, and the corresponding records are written back to the provenance layer.

## 3. Core modules of ALKEMIE Agent

Section 3 describes the module-level implementation of the architecture and bounded control loop introduced in Section 2. The first group of modules provides cross-cutting support for domain knowledge, context management, provenance recording, tool-calling iteration, and diagnosis. The second group provides functional capabilities for materials recommendation, AI-assisted structure modeling and editing, solver-facing execution, and active-learning-based self-iteration.

### 3.1 Knowledge base, RAG, and skill library

Domain expertise is distributed across the knowledge base, the retrieval-augmented generation (RAG) layer, and the skill library, with a distinct role assigned to each resource. The knowledge base stores software- and task-specific materials-computation entries; the RAG layer retrieves entries relevant to the active task context; and the skill library provides reusable procedural resources and access to permitted operations. This separation allows source knowledge, retrieval logic, and executable capability to be maintained independently while supporting a shared decision path from task interpretation to basic assistance or bounded tool-calling.

The knowledge base is organized around computational materials tasks rather than general conversational memory. Entries are indexed by software domain, calculation type, component identifier, parameter name, file type, parser output, diagnostic category, and structure-operation type. VASP entries cover input-file roles, convergence settings, output files, and common error patterns. GNEP entries cover structure sources, dataset preparation, training configuration, model files, and training-output interpretation. LAMMPS entries cover simulation-script components, force-field configuration, run-control parameters, thermodynamic output, trajectory files, and log inspection. AIMC entries cover exchange operations on the relevant sublattice and sampling parameters, including temperature, step number, and exchange attempts per step. Structure-modeling entries cover editing operations, format conversion, validation checks, and constraints on editable structure states. This organization provides software-specific context for parameter assistance, input preparation, parser interpretation, analysis guidance, and error-diagnosis assistance.

After the project context has been assembled, retrieval queries are formed from the task objective, detected software, component type, available structure or task metadata, parsed error messages, and human-specified constraints. Ranked entries from the relevant collections are then introduced into the reasoning and tool-selection context. Retrieved procedural resources do not enlarge the executable action space. Executable operations are exposed only through registered tools with predefined names, parameter schemas, and structured return types. A procedural resource can therefore guide task formulation, parameter selection, or diagnostic interpretation without changing the set of operations that the platform is permitted to perform.

The resulting request-to-action path connects the active project context, domain-specific retrieval, skill selection, permitted tool schemas, and execution-trace recording. Depending on the request, the same mechanism can produce explanations, suggest parameters, interpret parser outputs, guide analysis, or support error-diagnosis assistance. Retrieved entries ground software assistance, while physical assumptions, convergence behavior, and software-specific validity are assessed through solver outputs, validation checks, retained task records, and human review.

### 3.2 Context memory, task records, and database-supported provenance

Runtime context is separated from the persistent records used to reconstruct completed computational work. Short-term memory (project context) is assembled at the start of a request within the authenticated project and workspace boundary. It includes the project identity, recent interaction history, task objective/state, selected node, structure metadata, file references, intermediate observations, and pending confirmations. This context conditions routing, structure modeling, workflow design, parameter assistance, and analysis services, so that each operation is conditioned by the current computational situation rather than by an isolated instruction.

Persistent provenance is generated in parallel. Project metadata and task nodes are linked with structures, generated inputs, output files, logs, job status, status transitions, parser results, analysis artifacts, timestamps, and execution metadata. These records are treated as evidence of computational actions rather than as general memory. A reported output can therefore be traced to the corresponding node, files, execution state, and parsing process from which it is derived.

The execution trace connects transient context with persistent records by storing observable agent-mediated operations as ordered events. Text generation, status updates, tool requests, tool results, errors, and artifact references are normalized and associated with the relevant project, task node, file, job, or parsed result. The final response is consequently preserved together with the associated sequence of operations and observations.

Furthermore, an additional continuity mechanism is provided by optional cross-session project notes. Selected research summaries, human-confirmed observations, failure summaries, and lightweight references to project artifacts can be retained as project notes. These notes support later recall and reuse, whereas task records and database-supported provenance remain the authoritative reconstruction layer for calculations. Through this separation, subsequent requests can be informed by prior project state without weakening auditability, while calculation outputs, generated files, parsed results, and execution states remain anchored to task records and database-supported provenance.

### 3.3 Tool-calling iteration and error-diagnosis assistance

Coordination of route-specific actions is provided by a shared tool-calling runtime (Fig. 6). For each request, the objective, project context, active skill policy, permitted tool catalog, and intermediate state are assembled. The skill policy determines which knowledge resources and backend capabilities can be exposed, after which either a direct response is produced, or a registered tool is invoked. Task-route design, structure modeling, workspace operations, result analysis, and diagnostic assistance can therefore use the same control pattern while retaining task-specific context, adapters, and output schemas.

Each tool interaction is logged as a normalized runtime event. The event record may include visible text, tool requests, execution status, parser outputs, artifact references, structure updates, or error information. These events support real-time display, replay, debugging, and linkage to the provenance records described in Section 3.2. After each backend return, the observation is incorporated into the active context before another step is generated. Further tools may be invoked when additional evidence is required; otherwise, a final response or structured output is produced. When the configured iteration limit is reached, the available observations are consolidated without extending execution beyond the prescribed boundary. Multi-step reasoning over files, parsed results, retrieved records, and generated artifacts is thereby kept bounded and reviewable.

The same mechanism is used for error-diagnosis assistance. When abnormal parser results, tool observations, log entries, or status transitions are encountered, file evidence, parsed outputs, software-specific knowledge, and diagnostic rules are assembled as diagnostic context. The resulting response is returned as a reviewable likely cause and next-step suggestion. Subsequent correction or additional execution is performed only through a permitted tool call or software interface. This separation prevents diagnostic text from being treated as an executed recovery step.

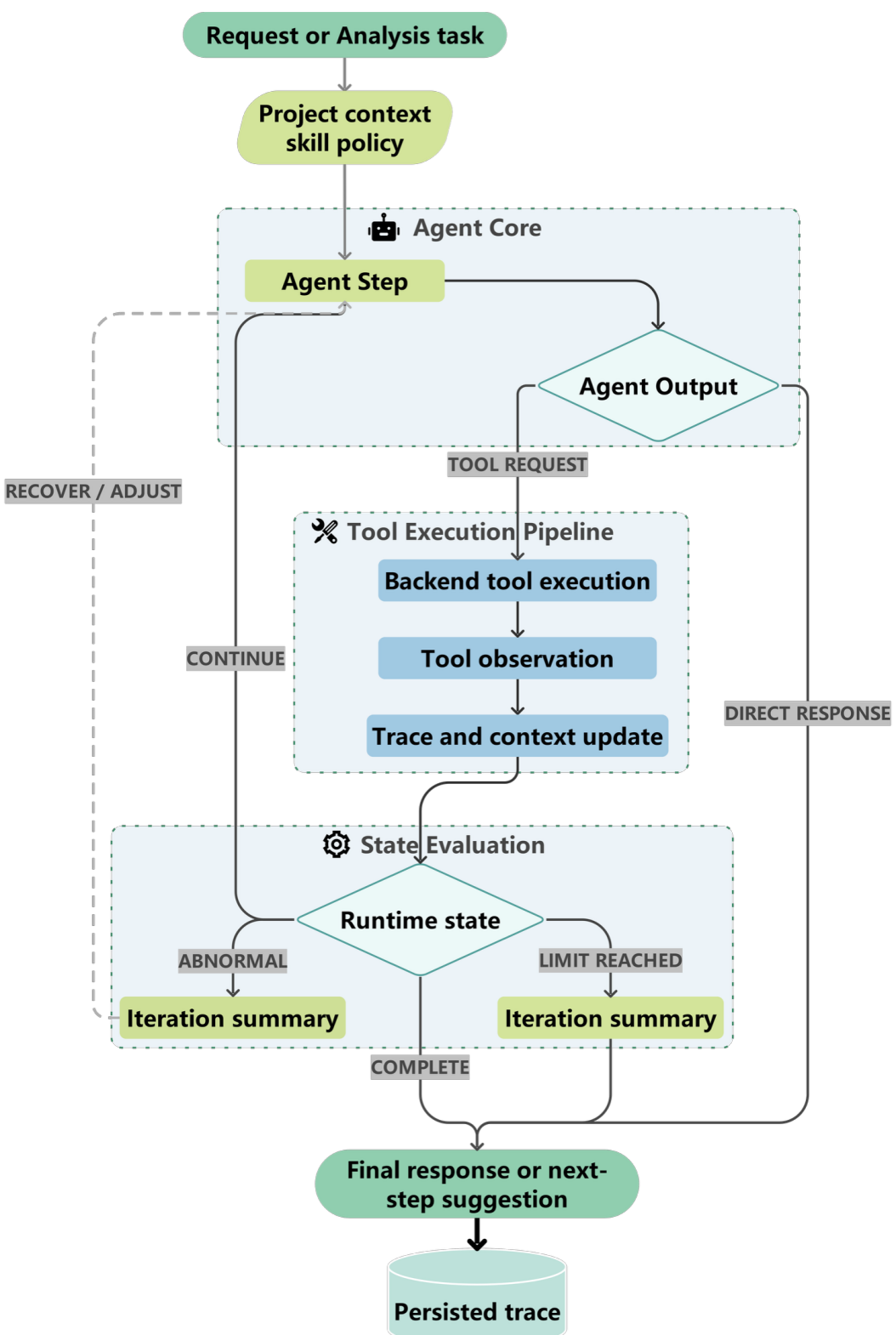

Fig. 6. Tool-calling iteration and error-diagnosis assistance in ALKEMIE Agent.

## 3.4 Materials recommendation module

The materials recommendation module provides an evidence-grounded route from design intent to human-reviewable candidate materials. At present, this route is implemented through a literature-derived knowledge graph rather than a free-form generative recommender. Web of Science (WOS) records are used as the primary input, with titles, abstracts, DOIs, journal names, and publication years retained as structured article metadata. Abstracts are then converted into graph records by an LLM-assisted extraction pipeline constrained by an explicit ontology. In this representation, articles, material systems, material variants, material properties, and physical mechanisms are connected by typed relations, so that reported materials, exhibited properties, explanatory mechanisms, and variant-to-parent dependencies can be preserved in a searchable form.

The construction pipeline and a representative graph structure are shown in Fig. 7. Field validation and document parsing are first applied to WOS tables. A unified parser checks titles, abstracts, DOIs, and fallback identifiers, and converts each publication into a standardized record for downstream extraction. Article-level extraction is then performed under a predefined schema, followed by post-processing of entity names and relationship endpoints. Duplicate or inconsistent material representations caused by differences in chemical-formula notation, subscript conventions, doping descriptions, or abbreviations are reduced through normalization. Material variants are retained as independent graph nodes and linked to their parent material systems, thereby preserving modification routes instead of merging them into a single material label.

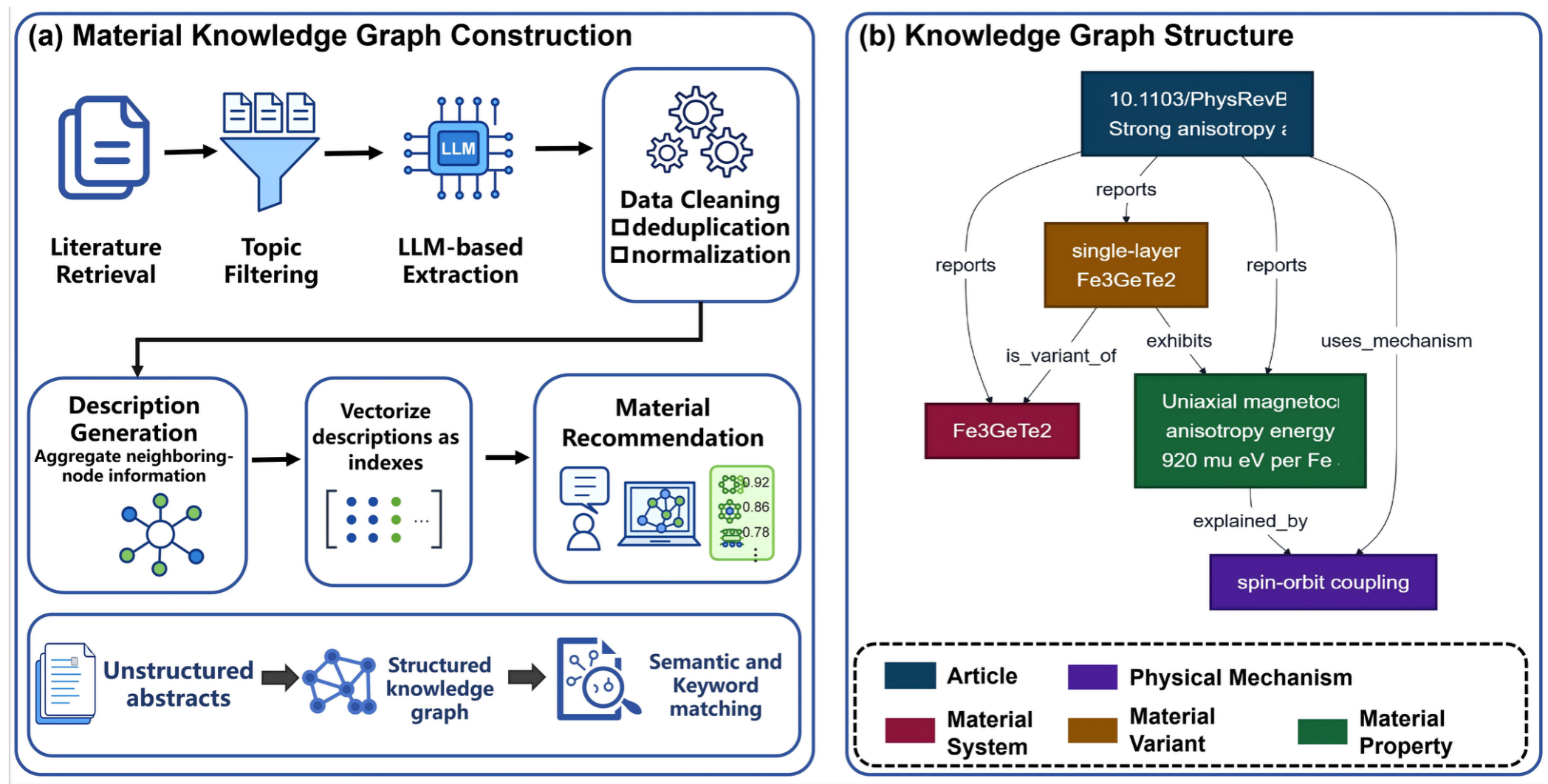

Fig. 7. Construction pipeline and representative structure of the knowledge graph using the case of two-dimensional magnetic-material. (a) Literature records are filtered, structured by LLM-assisted extraction, cleaned, enriched into material-centered descriptions, vectorized, and used for recommendation. (b) The representative graph connects articles, material systems, material variants, material properties, and physical mechanisms through typed relations.

Traceability is retained at both the entity and relationship levels. Source-publication metadata and original evidence snippets are kept with the extracted graph records, enabling each returned candidate to be linked back to the supporting papers and reported evidence. To make the graph directly usable for recommendation, material-centered descriptions are further generated by aggregating the properties, numerical values, mechanistic explanations, variants, and source publications associated with the same material family. These descriptions are embedded for semantic retrieval and are combined with recall based on material names, property terms, mechanism terms, and keywords.

When a recommendation query is submitted, the design target and constraints are mapped to the graph and the enriched retrieval space. For example, a request for two-dimensional magnetic materials that may retain ferromagnetic stability in few-layer form and show potential for a high Curie temperature can be handled by retrieving material nodes whose descriptions, properties, and evidence paths match these concepts. Candidate ranking is therefore determined by graph evidence and retrieval similarity rather than by generation alone. The output can include ranked materials, matched properties, evidence paths, source metadata, and a bounded evidence subgraph for inspection. In the current example of two-dimensional magnetic materials, the literature corpus contains 10,346 publications. After topic filtering and data extraction, the resulting material-centered graph contains 10,292 graph nodes and 230,417 typed edges, with publication metadata retained as source evidence.

### 3.5 AI-assisted structure modeling and editing

The AI-assisted structure modeling and editing module is designed for structure construction, modification, checking, conversion, and preparation for downstream calculations. A modeling request is interpreted in relation to the researcher's instruction, the currently loaded structure, recent project context, and relevant modeling knowledge. When an operation is successfully completed, an updated structure object is returned; when required information is missing or an operation fails, a human-reviewable response is generated instead. The module is therefore limited to structure preparation and editing within declared tool boundaries, while assessment of material stability, convergence, and application performance remains dependent on downstream calculations and human review.

The active structure is maintained as a state object containing composition, cell data, elements, coordinates, periodicity, source metadata, and derived information required for visualization, export, and downstream task connection. Natural-language instructions are mapped to schema-bound operations for cell and surface construction, species-level editing, coordinate or lattice modification, defect generation, conversion, validation, and symmetry analysis. As illustrated in Fig. 8, execution is delegated to backend structure services, and a revised payload is accepted only after successful completion.

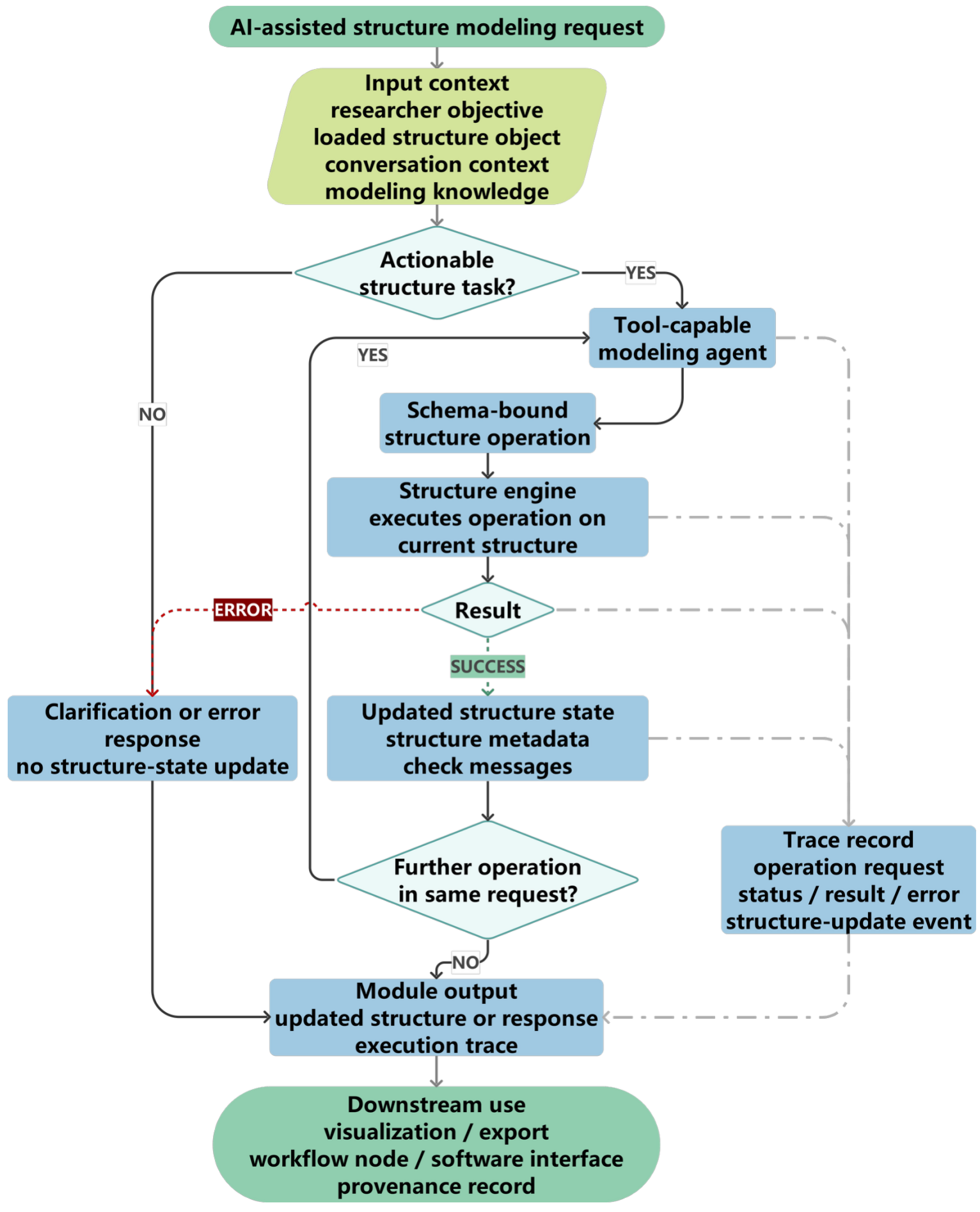

Fig. 8. Data flow for AI-assisted structure modeling and editing.

State updates are applied only after successful backend execution. For each operation, an ordered trace is generated to record the requested operation, execution status, returned result or error, and, when applicable, the updated structure payload. The updated structure state may be summarized and returned to the agent context for subsequent operations within the same request. Failed operations provide diagnostic information but do not replace the active structure state. Module outputs may include the updated structure, associated metadata, check messages, and execution trace. The updated structure can then be visualized, exported, saved as a reusable structure record, or passed to task nodes and simulation software interfaces when downstream connections are explicitly invoked.

### 3.6 Autonomous task execution and simulation software interfaces

Solver-facing execution is implemented through software-specific adapters coupled to a shared execution service (Fig. 9). Once a task objective or task node has been confirmed, the approved route is translated into an input package containing structure data, parameter settings, dependency information, execution metadata, and parser expectations. The adapter validates route-specific prerequisites, prepares the solver-readable inputs, and exposes a uniform interface for launch, status query, output discovery, and result extraction. This common contract allows different codes to be reached through the same execution service while keeping file formats, physical assumptions, and parser rules software-specific.

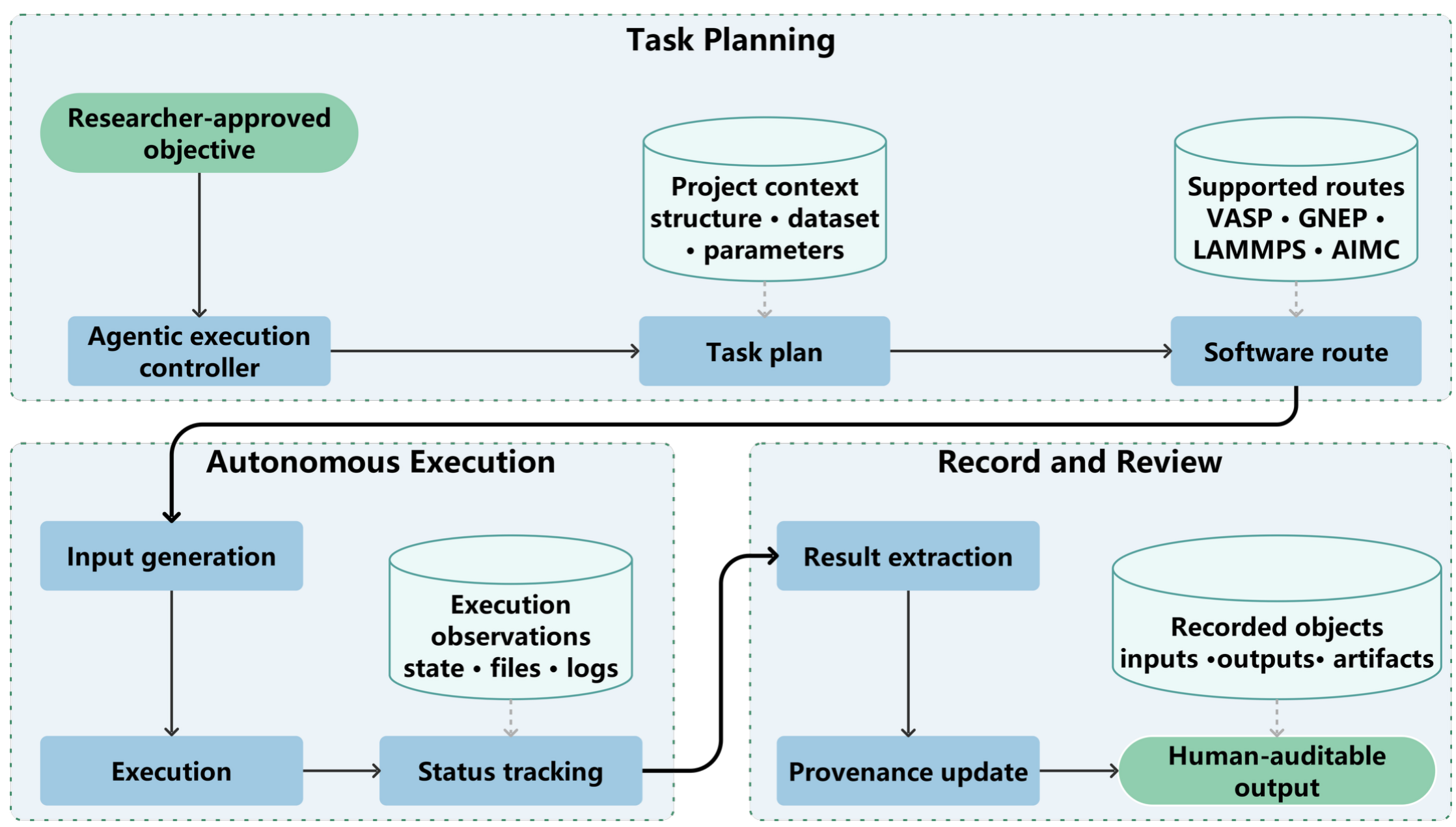

Fig. 9. Autonomous task execution interface for supported simulation software in ALKEMIE Agent.

The execution service separates route preparation from the deployment environment. Prepared tasks may be run through local execution, scheduler-mediated submission, or an authorized runtime service, depending on the configured installation. Regardless of backend, task states are updated through a common status model, and output availability is reported to the monitoring and parsing layer. Logs, recognized output files, generated artifacts, model files, and parsed quantities are then attached to the corresponding task record. This abstraction permits HPC-aware execution without embedding scheduler commands or deployment details into the agentic reasoning layer.

VASP, GNEP, LAMMPS, and AIMC are represented by distinct computational objects under this shared contract. VASP tasks are organized around structure-based electronic-structure calculations and parser outputs such as energies, forces, phonon data, electronic quantities, or convergence records, depending on the selected calculation type. GNEP tasks connect structure sources and first-principles labels with dataset construction, training configuration, model artifacts, and evaluation outputs. LAMMPS tasks bind structures or data files to force-field definitions, simulation controls, thermodynamic logs, trajectories, and post-processing results. AIMC tasks associate initial structure, MC sampling parameters with sampled equilibrium configurations within MC trajectory. For each task, the generated inputs, execution metadata, detected outputs, parsed records, and execution trace are linked, thereby allowing subsequent interpretation to be performed with reference to the computational conditions under which the result is produced.

### 3.7 Active-learning-based self-iteration for materials screening

The active-learning module in ALKEMIE is implemented as a property-driven screening loop for reducing the number of labeled evaluations required during candidate search. Composition and structure descriptors are generated with Matminer[35], after which an initial labeled subset is sampled; by default, 5% of the structures are used for initial training and the remaining entries constitute the candidate pool. Optional preprocessing, including data cleaning, feature scaling, variance filtering, and SelectKBest feature selection, can be included in the training pipeline. Regression is currently supported with Gaussian process regression (GPR) and XGBoost[36][36], and model quality is estimated through user-defined k-fold cross-validation. The iterative data-model-selection loop is summarized in Fig. 10.

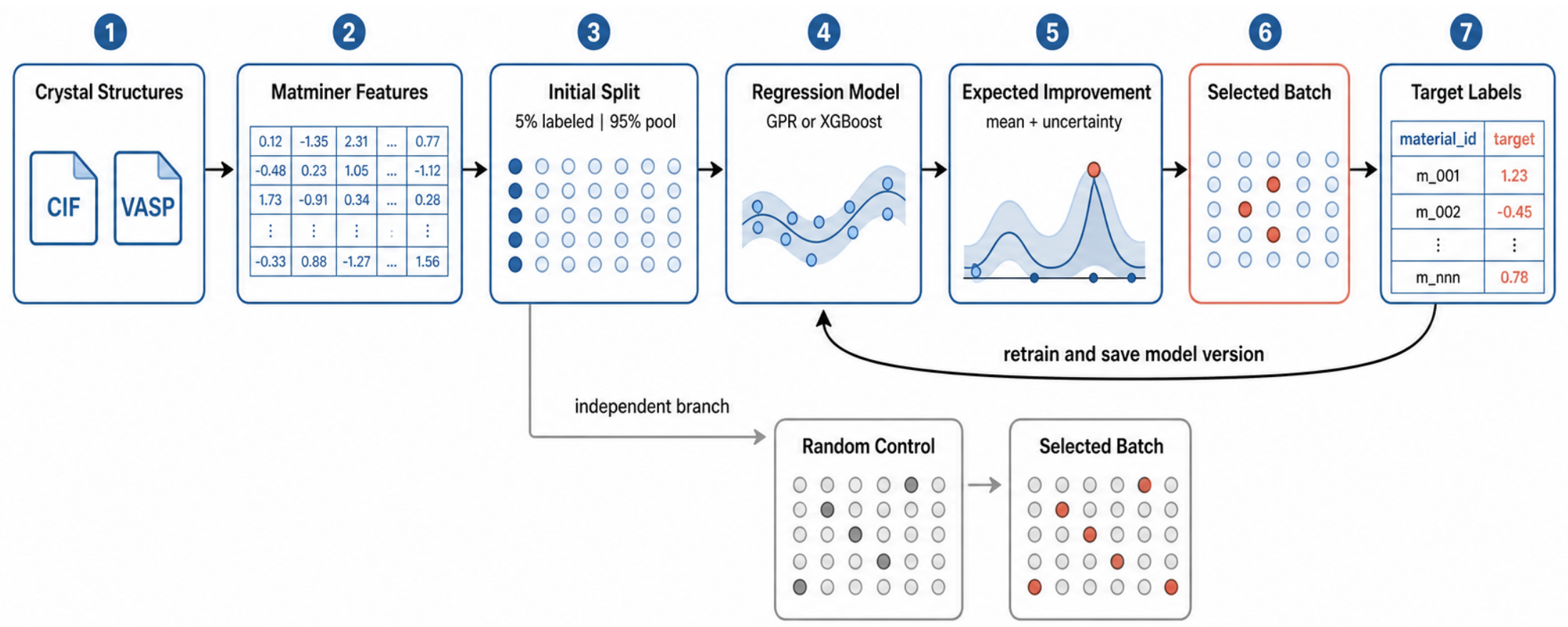

Fig. 10. Active-learning loop implemented in ALKEMIE Agent. Crystal structures are converted into numerical descriptors and divided into an initial labeled set and an unlabeled candidate pool. A regression model predicts the target and its uncertainty. Expected improvement ranks the candidates, and the selected structures are returned for labeling. The new labels are added to the training set before the next model is built.

Candidate prioritization is based on expected improvement (EI), which combines exploitation of a favorable predicted target value with exploration of regions where predictive uncertainty remains high[37,38]. For maximization or minimization, the improvement term is written as:

$$I(x) = s[\mu(x) - f_{best}] - \xi \quad (1)$$

$$EI(x) = I(x)\Phi\left(\frac{I(x)}{\sigma(x)}\right) + \sigma(x)\phi\left(\frac{I(x)}{\sigma(x)}\right) \quad (2)$$

where s=1 for maximization and s=-1 for minimization. Here, μ and σ are the predicted mean and uncertainty, $f_{best}$ is the best observed value, and ξ controls exploration. In this implementation, uncertainty is obtained directly from GPR, whereas an empirical uncertainty proxy for XGBoost is estimated from cross-validation residuals. The functions Φ and φ denote the cumulative distribution function and probability density function of the standard normal distribution, respectively. After ranking, the highest-scoring candidates are exported together with their features. Once new target values are supplied and the labeled set is updated, then the model is re-trained, and the next selection round can be initiated. A random-selection branch may be maintained in parallel as a control, with its data, models, and selected structures kept separate from the EI branch. In this way, ALKEMIE Agent realizes the central active-learning objective in materials screening: candidate selection is biased toward high-value regions of the search space while the number of additional labeled evaluations is reduced[39].

## 4. Representative applications and computational demonstrations

Representative cases are examined to illustrate how the modules described in Section 3 can be combined across the materials-design task chain. The demonstrations cover knowledge-graph-based recommendation, structure preparation, first-principles calculation, interatomic-potential training, molecular dynamic simulation, atomic-configuration sampling, and active-learning screening. Assessment is based on retained structures, task states, calculation files, parsed results, and execution records. In the following examples, the LLM request is accomplished by DeepSeek-V4-flash.

### 4.1 Materials recommendation for 2D magnetic systems

Materials recommendation is examined using the two-dimensional magnetic-materials knowledge graph. Firstly, a request targeting the van der Waals two-dimensional ferromagnets with high Curie temperature and strong magnetic anisotropy is submitted through the user interface (Fig. 11). The recommendation module retrieves $CrI_3$, $Fe_3GeTe_2$, and $VS_2$ as seed materials that already have strong graph evidence for the target. Around each seed material, the graph then separates into two types of recommendation results. The first type consists of known variants of the seed material, such as $CrBr_3/CrI_3$ mixed bilayer, AFM bilayer $CrI_3$, and $Ag(CrI_3)_4$ for $CrI_3$, which summarize reported modification strategies including composition mixing, layer-number control, and intercalation or chemical modification. The second type consists of graph-expanded candidate materials, including $CrWCl_6$, $VI_3$, $CrSe_2$, $Cr_2Ge_2Te_6$, $FeCl_2$, and $CrSiTe_3$. These candidates are not treated as verified variants of the seed materials; instead, they are connected to the target profile through shared properties, mechanisms, or evidence paths, for example, shared ferromagnetic behavior or Curie-temperature-related properties.

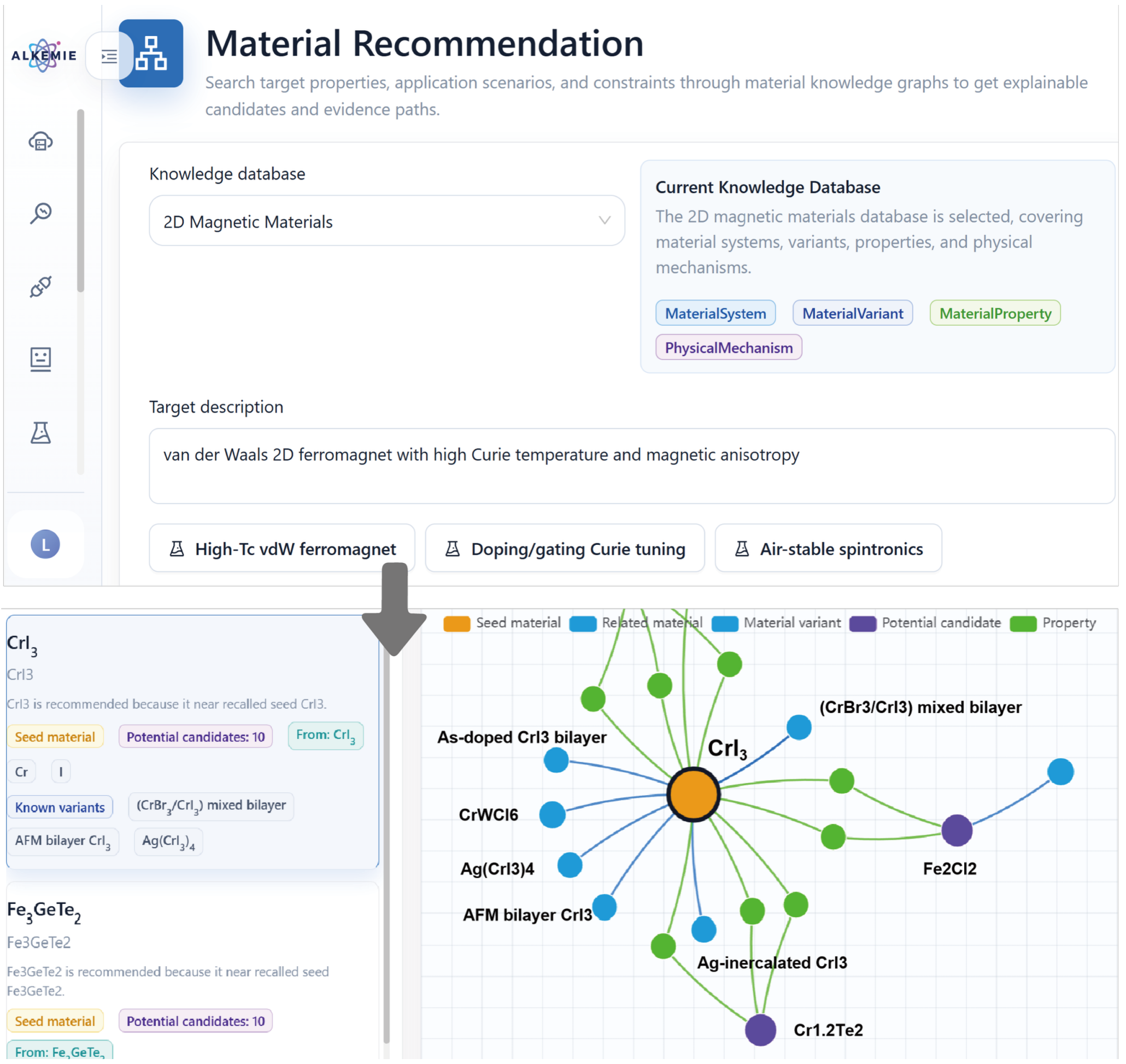

Fig. 11. Materials recommendation based on the two-dimensional magnetic-materials knowledge graph. A target profile for a high-Curie-temperature van der Waals ferromagnet with magnetic anisotropy is connected to seed materials, literature-reported variants, graph-expanded candidates, properties, and supporting evidence paths.

The separation between reported variants and graph-expanded candidates provides both interpretability and design-transfer value. Reported seed variants retain literature-supported modification routes, including composition mixing, layer-number control, intercalation, chemical modification, strain engineering, electrostatic gating, and heterostructure construction. Graph-expanded candidates, in contrast, are linked to the target profile through shared properties, mechanisms, or evidence paths, and therefore serve as evidence-linked candidate hypotheses rather than ungrounded generations. In this form, the recommendation record connects candidate prioritization with traceable design rationales: modification strategies observed around seed materials can be transferred as reference routes for subsequent calculation/experiment validation.

### 4.2 Computation-ready structure modeling

Atomic structure modeling is the basis for most of the calculations, here this function is illustrated based on a 16-atom $Cu_3Sn$ structure in the $Fm\bar{3}m$ space group (Fig. 12a). The structure modeling request comprises construction of a $2 \times 2 \times 2$ supercell, substitution of one Cu site by Au, and generation of a (100) slab with a 15 Å vacuum region. The structure views and execution trace are summarized in Fig. 12b, c.

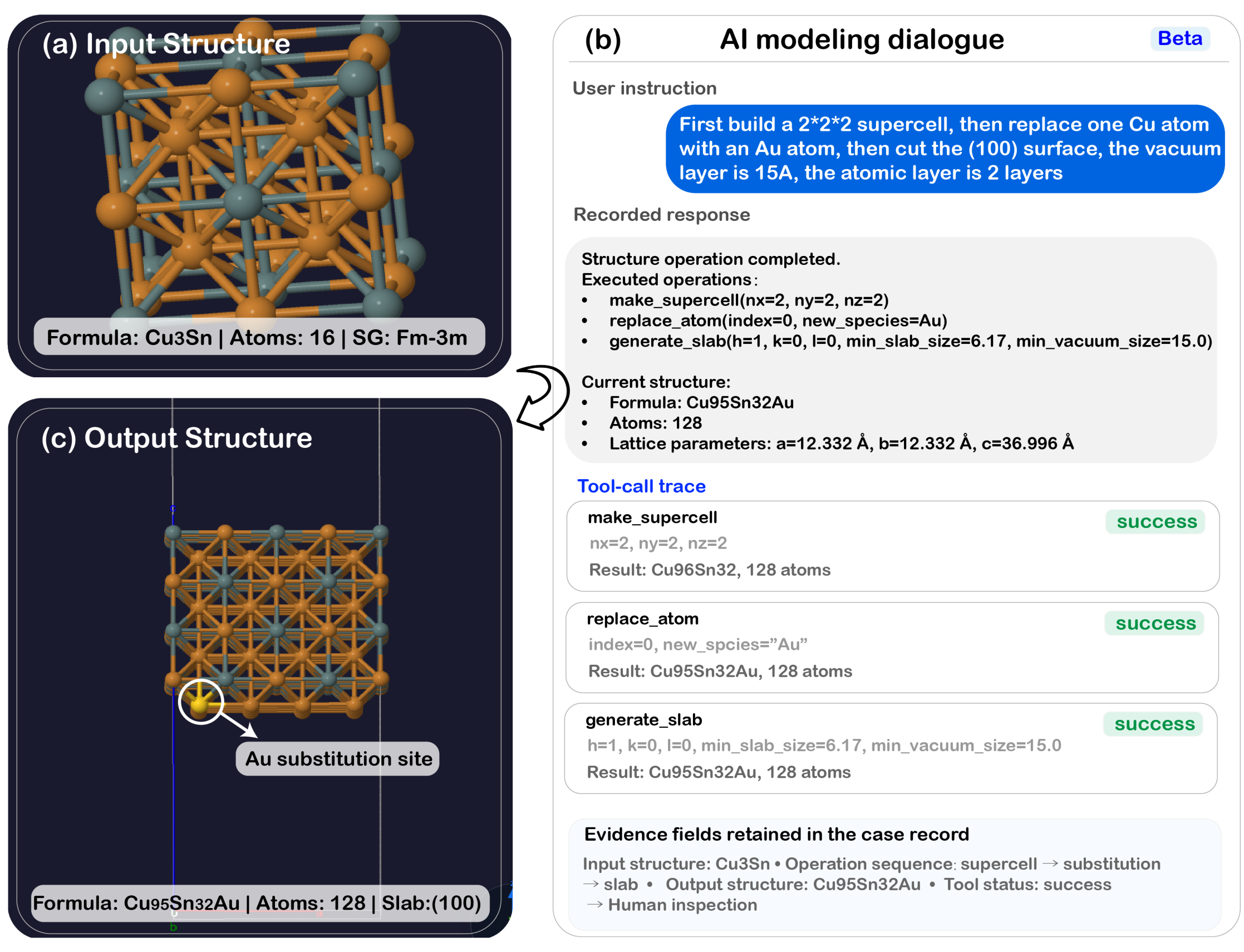

Fig. 12. AI-assisted structure modeling. (a) Input $Cu_3Sn$ structure containing 16 atoms in the $Fm\bar{3}m$ space group. (b) Recorded transformation sequence and execution statuses for supercell construction, Cu-to-Au substitution, and (100) slab generation. (c) Resulting 128-atom $Cu_{95}Sn_{32}Au$ slab with the substituted Au site indicated. A minimum slab thickness of 6.17 Å and a vacuum thickness of 15.0 Å are applied.

The structure building and transformation are completed through three sequential tool calls. A 128-atom $Cu_{96}Sn_{32}$ supercell is first generated using "make_supercell" (nx=2, ny=2, nz=2). The Cu atom at index 0 is subsequently replaced by Au using "replace_atom" (index=0, new_species=Au), after which the composition is changed to $Cu_{95}Sn_{32}Au$. The slab geometry is then generated using "generate_slab" (h=1, k=0, l=0, min_slab_size=6.17, min_vacuum_size=15.0). A success status is returned for each tool call. The resulting structure is generated with lattice parameters $a = b = 12.332$Å and $c = 36.996$Å as in Fig. 12c.

This case demonstrates that ALKEMIE Agent can translate a natural-language modeling objective into a coordinated sequence of structure operations, including supercell construction, species-level substitution, slab generation, structure-state updating, and visualization-ready output generation. The input structure, operation sequence, tool parameters, execution status, output composition, atom count, lattice metadata, and inspected structure are retained in the case record, thereby linking the generated geometry to a reproducible modeling trace. The resulting structure can therefore be used as a solver-ready starting structure for subsequent relaxation, surface-energy evaluation, or task-specific simulation after human inspection. In this arrangement, AI-assisted structure modeling and editing functions as a traceable preparation layer that connects conversational structure design with downstream VASP, LAMMPS, GNEP, AIMC, or other solver-specific routes.

### 4.3 Phonon spectrum calculation by VASP and Phonopy

The first solver-based case involves a finite-displacement phonon calculation for monolayer 2H-$MoS_2$. The calculation is initiated by constructing the monolayer model through an LLM-enabled pymatgen skill, followed by visual inspection of the generated structure (Fig. 13a, b). The calculation workflow comprises structural relaxation, finite-displacement force calculations, and phonon post-processing. Following human confirmation, these stages are instantiated as connected task nodes, and their execution states are exposed in the execution view (Fig. 13c, d).

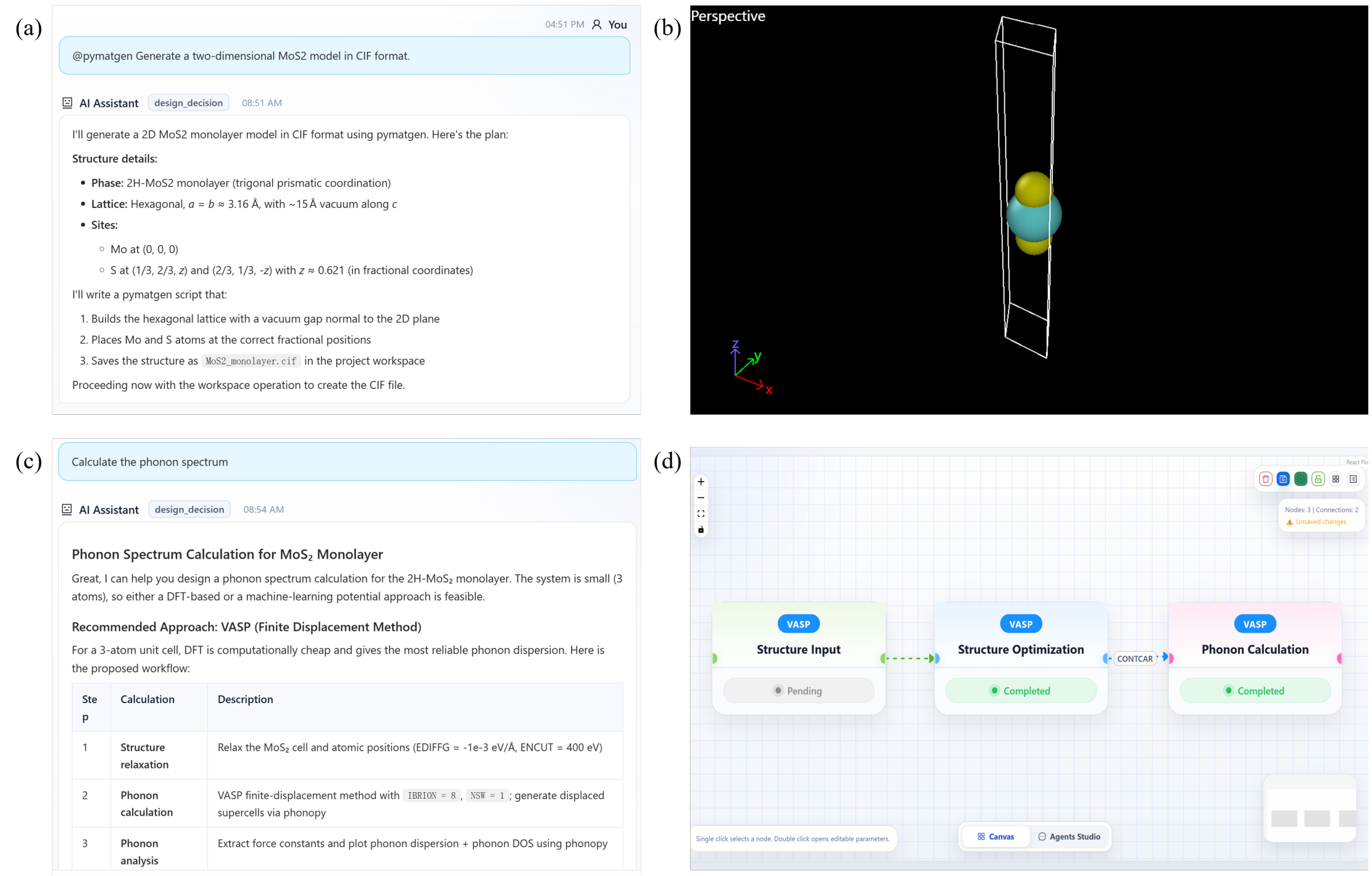

Fig. 13. Task specification and task-graph construction for the monolayer 2H-$MoS_2$ phonon calculation. (a) Specification and generation of the monolayer structure. (b) Three-dimensional view of the generated model. (c) Selection of the finite-displacement calculation strategy. (d) Instantiated workflow containing structure input, structure optimization, and phonon-calculation nodes.

The Phonopy package[40] is employed to generate displaced supercells and compute atomic forces, and each displaced supercell is evaluated as an independent scheduler-managed task with a shared upstream dependency. Completion and convergence are verified from the corresponding outputs before the forces are assembled for phonon-band and DOS calculations. The parsed record includes the frequency range, number of branches, high-symmetry labels, imaginary-mode check, and dynamical-stability assessment. Any failed displacement is propagated to the composite task state together with the number of affected tasks.

Traceability is maintained by associating the input parameters and intermediate decisions with the raw outputs, stage-level states, and per-displacement status records. The result record is also linked to the dimensionality assessment used during task-route construction. This association preserves the provenance chain from initial model preparation to the final stability assessment and allows the post-processing step to be reproduced from the retained calculation artifacts. The calculated phonon dispersion along Γ–M–K–Γ and the total phonon DOS are shown in Fig. 14, which is automatically generated by the platform.

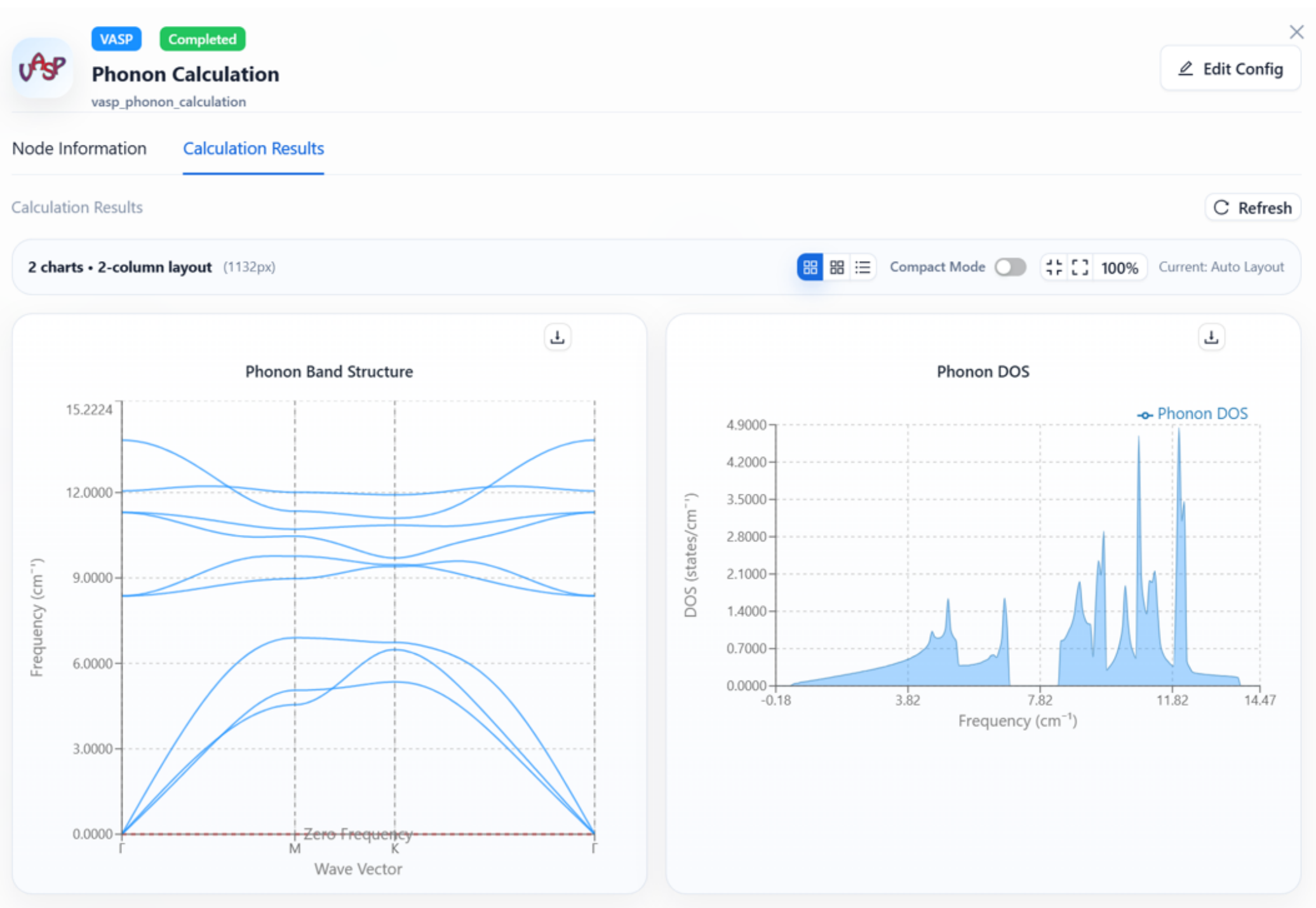

Fig. 14. Parsed outputs from the monolayer 2H-$MoS_2$ phonon calculation: phonon dispersion along Γ–M–K–Γ (left) and total phonon density of states (right).

ALKEMIE Agent also supports additional domain-oriented interpretation. As shown in Fig. 15, the Analysis agent is invoked to provide reviewable interpretation of the phonon results, including the imaginary-mode check, characteristic vibrational modes, frequency-gap discussion, and possible follow-up calculations such as thermal-conductivity or free-energy evaluations. These interpretations are retained as suggestions for researcher review.

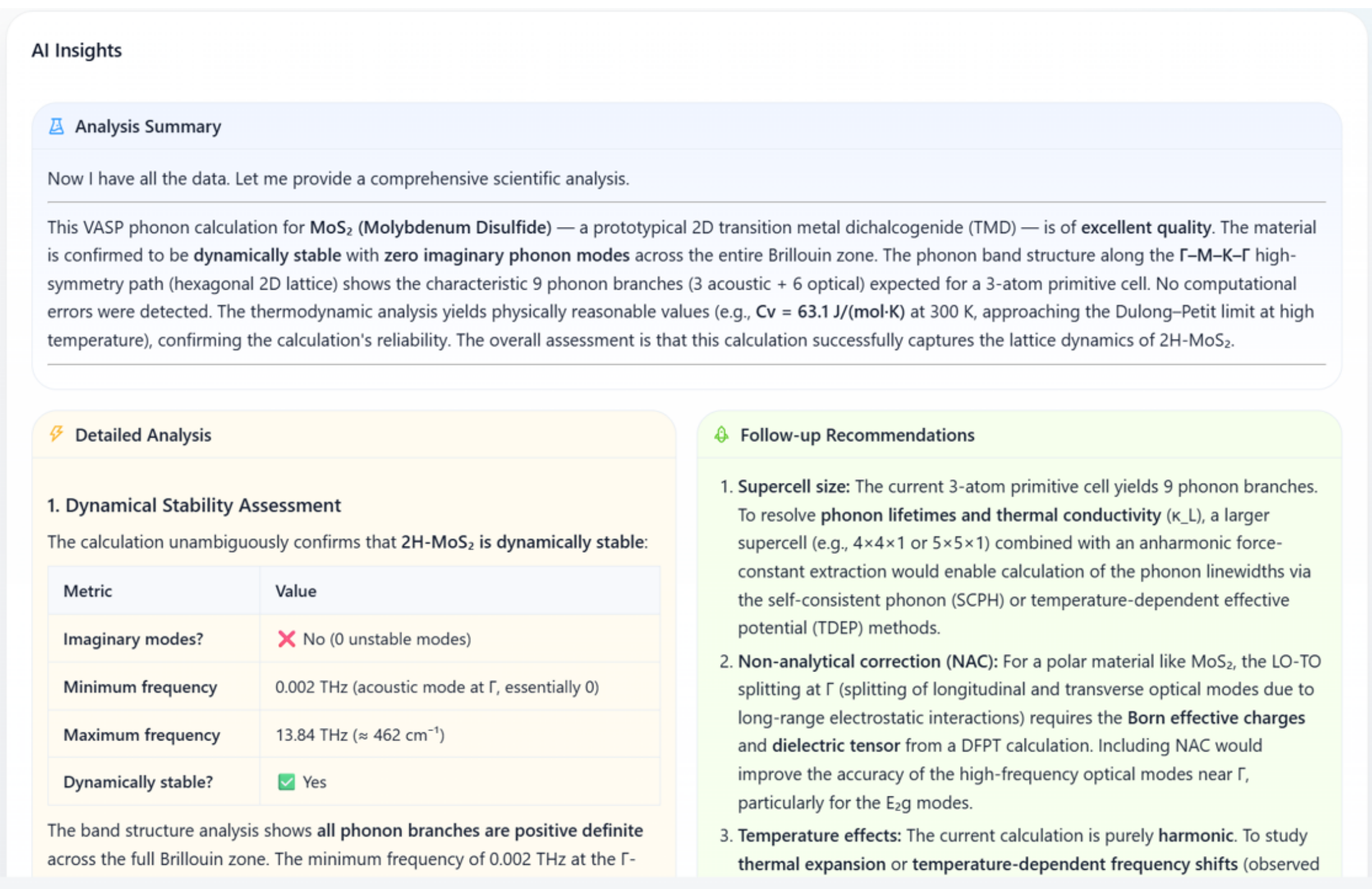

Fig. 15. AI-assisted interpretation of the parsed phonon results, including the imaginary-mode check, frequency summary, and human-approved next-step suggestions.

### 4.4 MLP training via GNEP

Following the same task-graph construction used in the above case, a representative route of MLP training via GNEP is assembled for Ag-Cu-Sn system. As shown in Fig. 16, following human confirmation, the requested calculation is decomposed into six stages: original structure ingestion, derived structure generation, VASP labeling, dataset build, model training, and model evaluation. The corresponding proposal contains 6 nodes and 5 directed dependencies, and is exposed through preview, comparison, and application controls before the insertion into the task canvas.

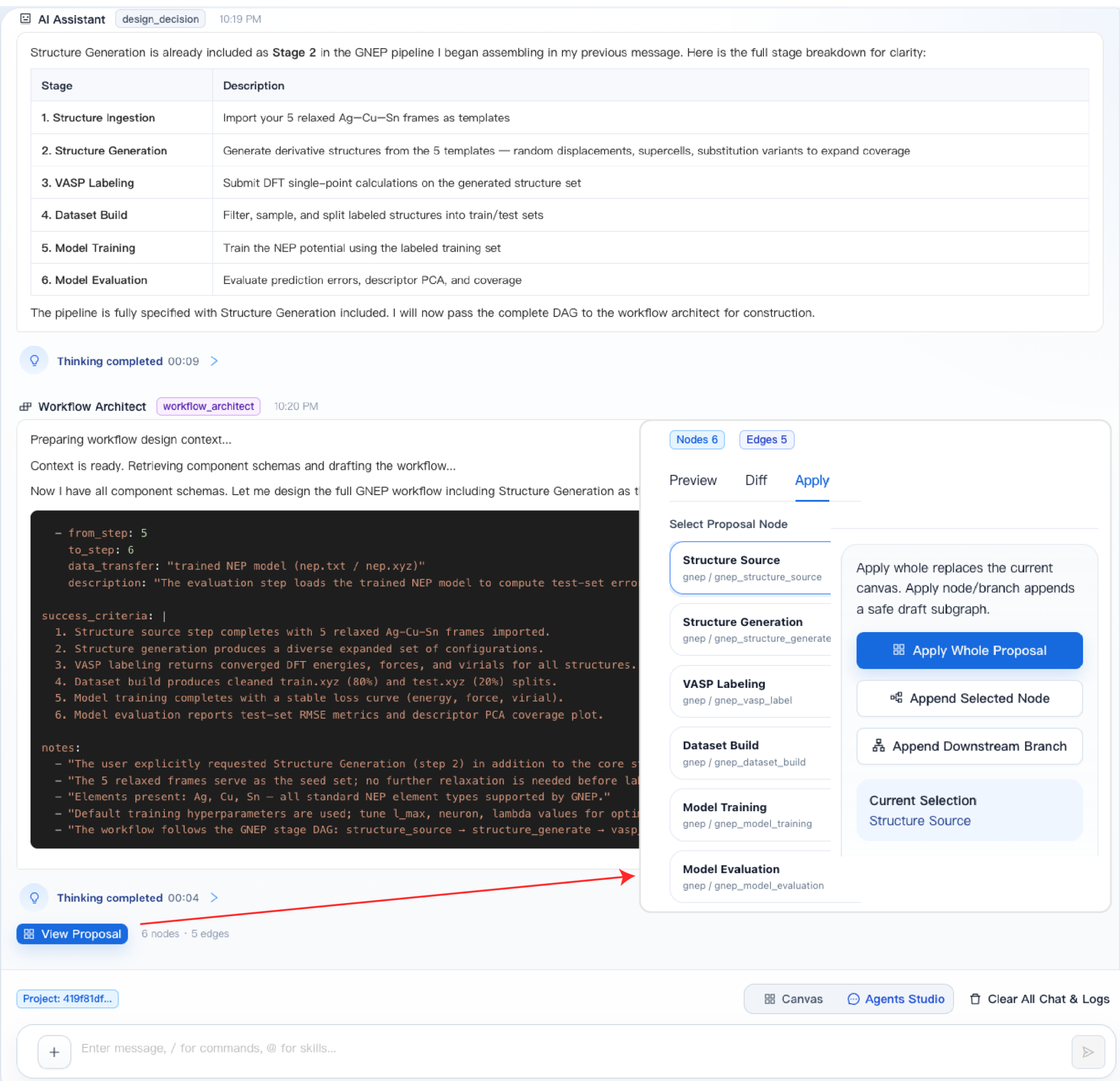

Fig. 16. Human-supervised formulation of a GNEP task in ALKEMIE Agent. A six-stage route containing structure ingestion, structure generation, VASP labeling, dataset construction, model training, and model evaluation is presented as a six-node, five-edge proposal for review and application.

Following confirmation, the proposed route is instantiated as a node-based task graph (Fig. 17a). Structure artifacts and associated metadata are transferred from structure ingestion to generation and labeling. The labeled configurations are subsequently passed to dataset construction, where training and test subsets are prepared for model fitting. The trained model artifact and associated runtime records are then transferred to the evaluation stage. Completed states are displayed for all 6 nodes, providing a traceable association among task topology, inter-stage artifacts, and recorded execution status.

Upstream candidate preparation is configured through the Structure Recipe

Workbench shown in Fig. 17b. Multiple transformation branches are derived from the source structures, including random perturbation, random doping, random slab generation, cell scaling, supercell construction, vacancy-defect construction, and interstitial-defect construction. The displayed workbench record contains 5 source structures, 7 recipe branches, 428 final structures, and two dropped structures. The generated structure set is positioned upstream of VASP labeling and subsequent dataset construction.

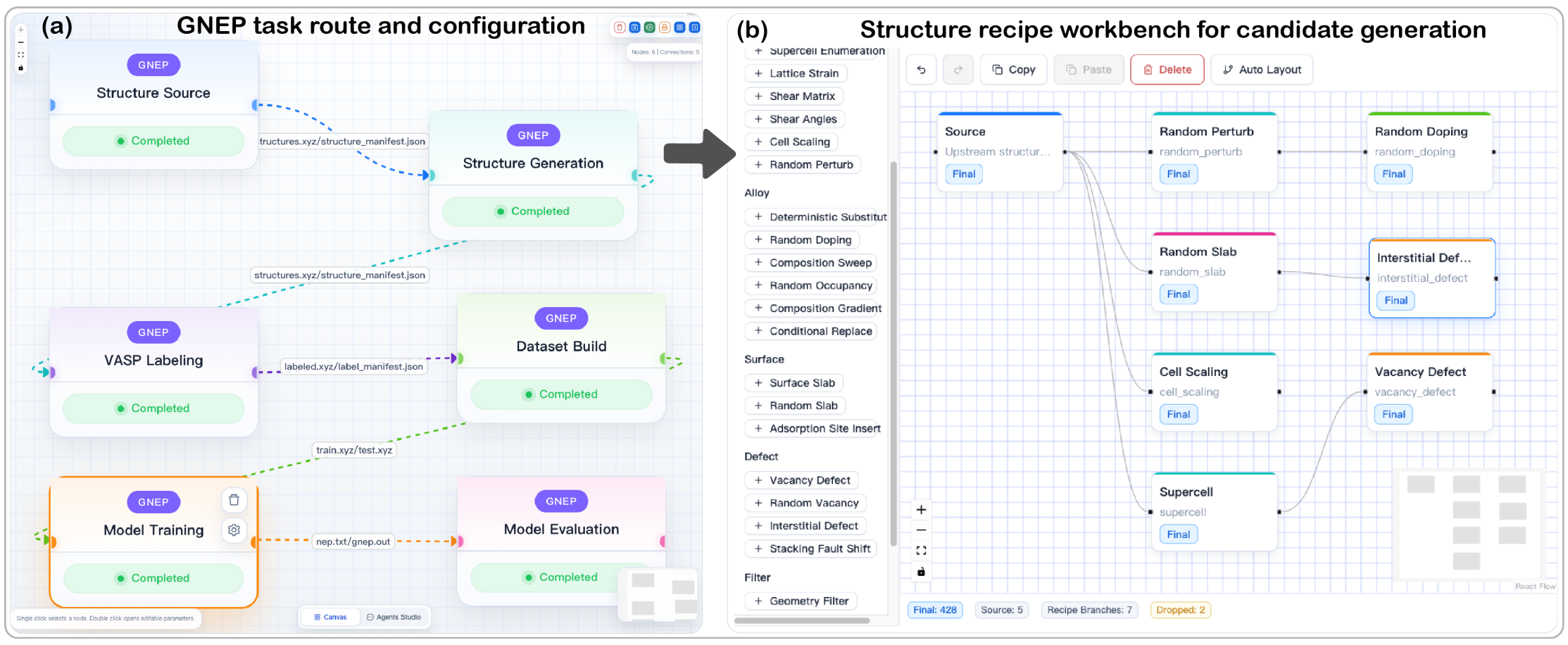

Fig. 17. Instantiated GNEP route and upstream structure preparation. (a) Node-based GNEP task graph with recorded completion states and explicit inter-stage artifact transfer. (b) Structure Recipe Workbench used to expand five source structures through seven transformation branches, resulting in 428 final structures and two dropped structures.

This case establishes traceable task decomposition, route review, task-graph instantiation, structure-recipe configuration, and execution-state recording; potential transferability can be assessed when parsed training and evaluation metrics are included. Parsed training and evaluation metrics, including error and coverage analyses, will extend this demonstration from workflow traceability to potential-quality assessment.

### 4.5 Uniaxial tension evaluation of high-entropy alloy via LAMMPS

In ALKEMIE Agent, molecular dynamics simulations through LAMMPS are represented by task templates whose current implementation covers system setup, energy minimization, thermal equilibration, uniaxial loading, elastic-constant evaluation, and Monte Carlo sampling. As illustrated in Fig. 18, the System Setup module serves as the entry point for task construction, where structural models, force fields, and simulation parameters are initialized. Depending on the computational objective, downstream modules can be proposed and connected after agent interpretation and human confirmation. For example, an elastic-property calculation route can be constructed by combining the System Setup and Elastic Constants components, whereas a mechanical-deformation route further incorporates Energy Minimization, Thermal Equilibration, and Uniaxial Tension components.

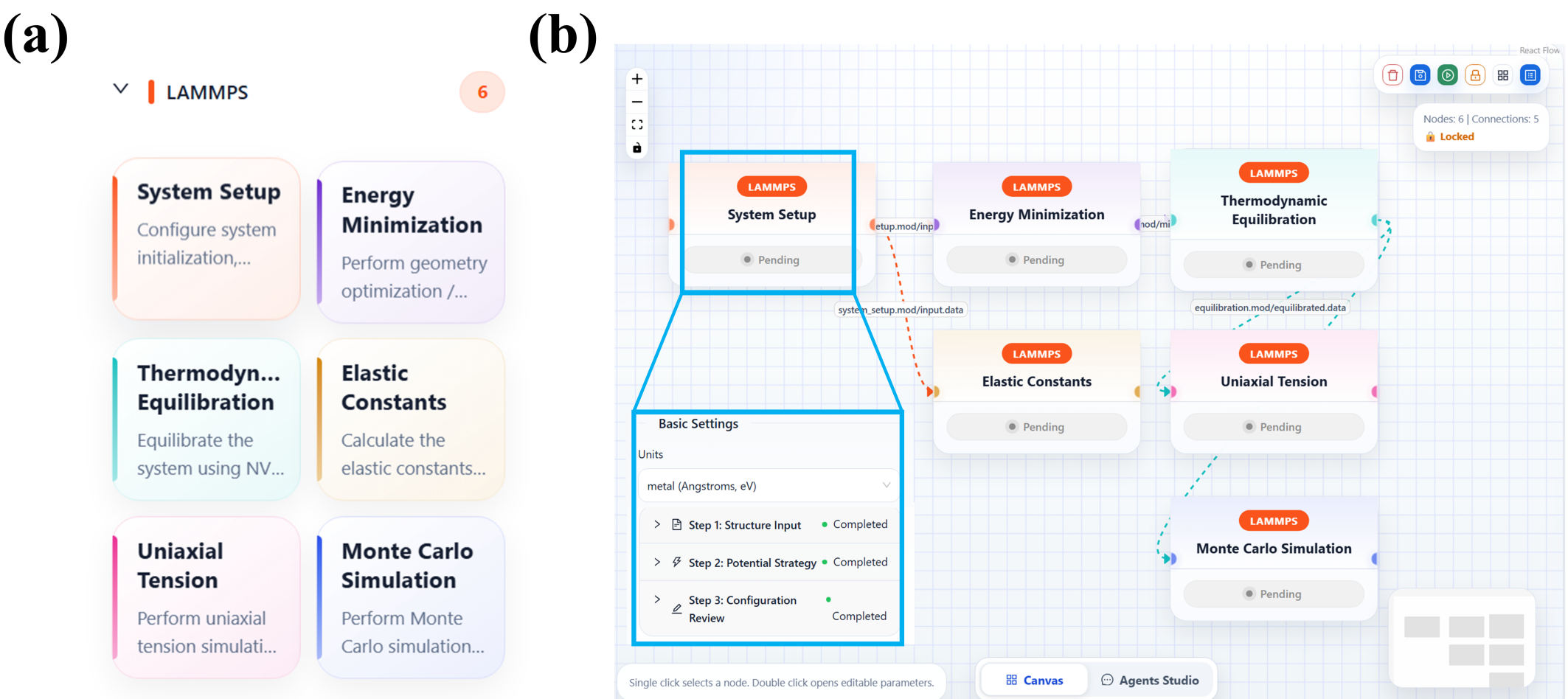

Fig. 18. Molecular dynamics simulation task-graph via LAMMPS. (a) Reusable simulation components. (b) Task-orchestration canvas showing component connectivity and execution order. The expanded System Setup component illustrates the configuration process, including structure input, potential strategy, and configuration review prior to workflow execution.

A uniaxial tension simulation for the TiZrVNb high-entropy alloy (HEA) supercell with a length scale of approximately 6.6 nm is selected as a representative example. As shown in Fig. 19a, after the request “Write a LAMMPS task for computing the uniaxial tension of the TiZrVNb high-entropy alloy.”, the agent first interprets the scientific objective and formulates a computational strategy. Based on domain knowledge and route-planning rules, the agent decomposes the task into four stages: structure construction, potential selection, thermodynamic equilibration, and uniaxial deformation (Fig. 19c). The TiZrVNb alloy is represented as a body-centered-cubic (BCC) high-entropy solid solution. A statistically sampled random solid-solution supercell is constructed (Fig. 19b) to approximate chemical disorder in the HEA model.

For interatomic interactions, a machine-learning-based NEP potential is adopted to accurately capture anharmonic deformation and defect evolution under large strain. Prior to mechanical loading, the system is equilibrated to remove residual stresses and achieve thermodynamic stability. Subsequently, uniaxial tension is applied along a prescribed crystallographic direction under a constant engineering strain rate, while lateral directions are controlled to maintain zero transverse stress conditions. During deformation, virial stress and strain are continuously recorded to construct the stress-strain response.

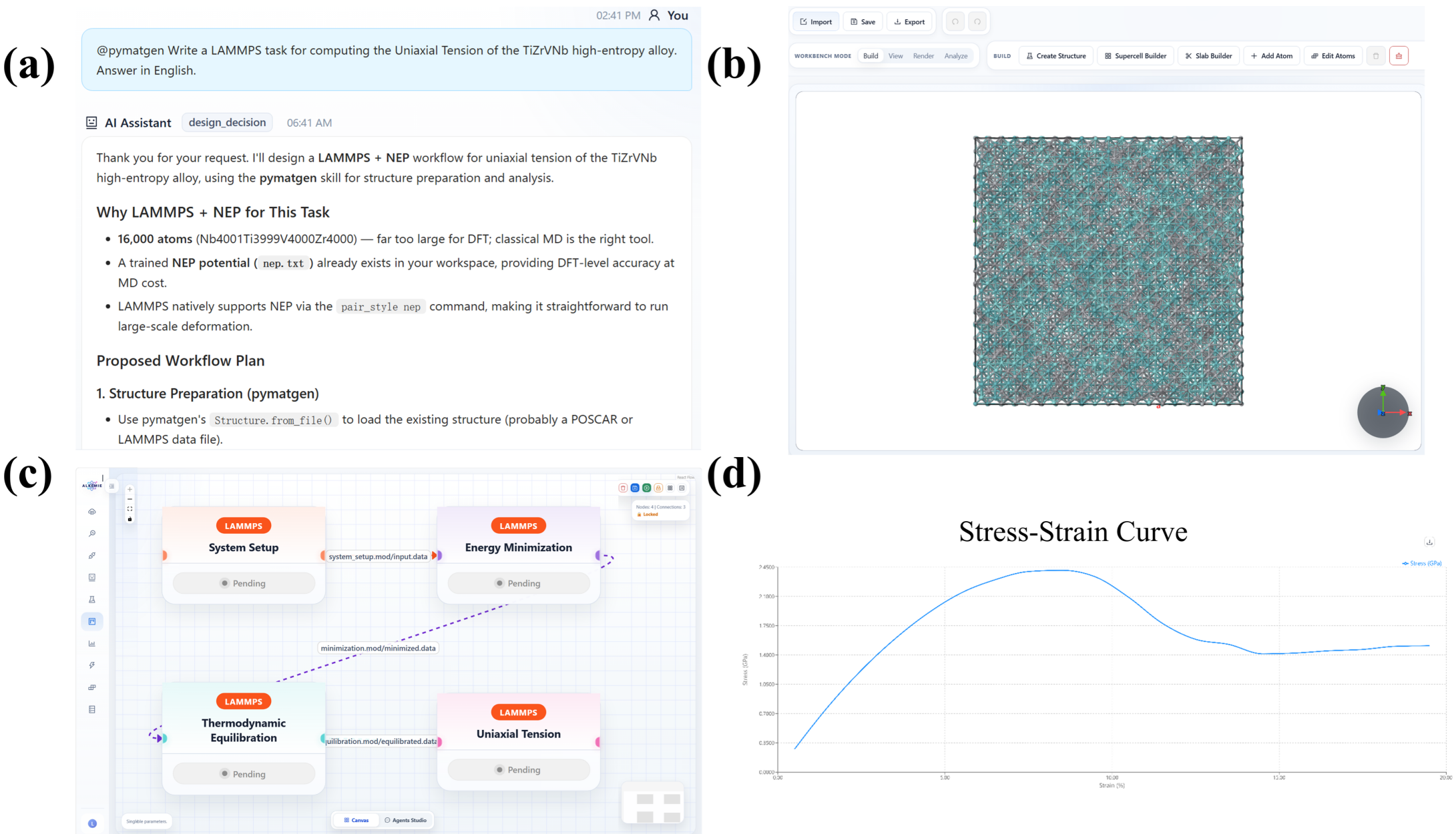

Fig. 19. LAMMPS task-graph for uniaxial tension simulation of the TiZrVNb high-entropy alloy. (a) Calculation specification. (b) Generated body-centered-cubic random-solid-solution supercell. (c) Executed workflow linking system setup and energy minimization, thermodynamic equilibration, and uniaxial tension evaluation. (d) Stress–strain curve from simulation results.

After the simulation, as illustrated in Fig. 19d, a complete stress-strain curve is generated, from which key mechanical properties such as yield strength and ultimate tensile strength can be identified. All results are stored with metadata including loading direction, strain rate, temperature, and potential model, supporting reproducibility and provenance tracking.

### 4.6 Atomic arrangement sampling via AIMC

In multicomponent materials, atomic arrangements strongly influence thermodynamic stability and physical properties. Conventional special quasi-random structures (SQS)[41] provide an efficient approximation for chemically disordered alloys, but they may not capture the short-range ordering effects commonly present in real materials. Therefore, ALKEMIE Agent incorporates AIMC to sample configurational space at a specified temperature and identify energetically favorable atomic arrangements[42–45].

By integrating structure manipulation, energy calculation, and trajectory analysis, this module supports exploration of low-energy atomic arrangements in multicomponent materials. As illustrated in Fig. 20, ALKEMIE Agent supports the construction and execution of AIMC tasks for multicomponent materials. The quaternary alloy $(Re_{0.5}Nb_{0.5})(S_{0.5}Se_{0.5})_2$ is employed as the representative example, which contains two independent sublattices. In the metal sublattice, Re and Nb atoms can be exchanged, whereas in the chalcogen sublattice, S and Se atoms can be swapped. During the MC simulation, these exchange operations are repeatedly performed to sample the configurational landscape and identify energetically favorable arrangements.

The convergence and sampling behavior of the MC simulation are summarized in Fig. 20. The left panel shows the autocorrelation function of the sampled configurations. The autocorrelation decreases below 1/e under the selected criterion, suggesting that the sampled configurations are effectively decorrelated within the analyzed trajectory. The right panel displays the accepted and rejected atomic-exchange trials during the MC trajectory, providing a direct representation of the configurational sampling process governed by the Metropolis criterion[45]. The low-energy arrangement obtained by the AIMC module can be employed for downstream property calculations.

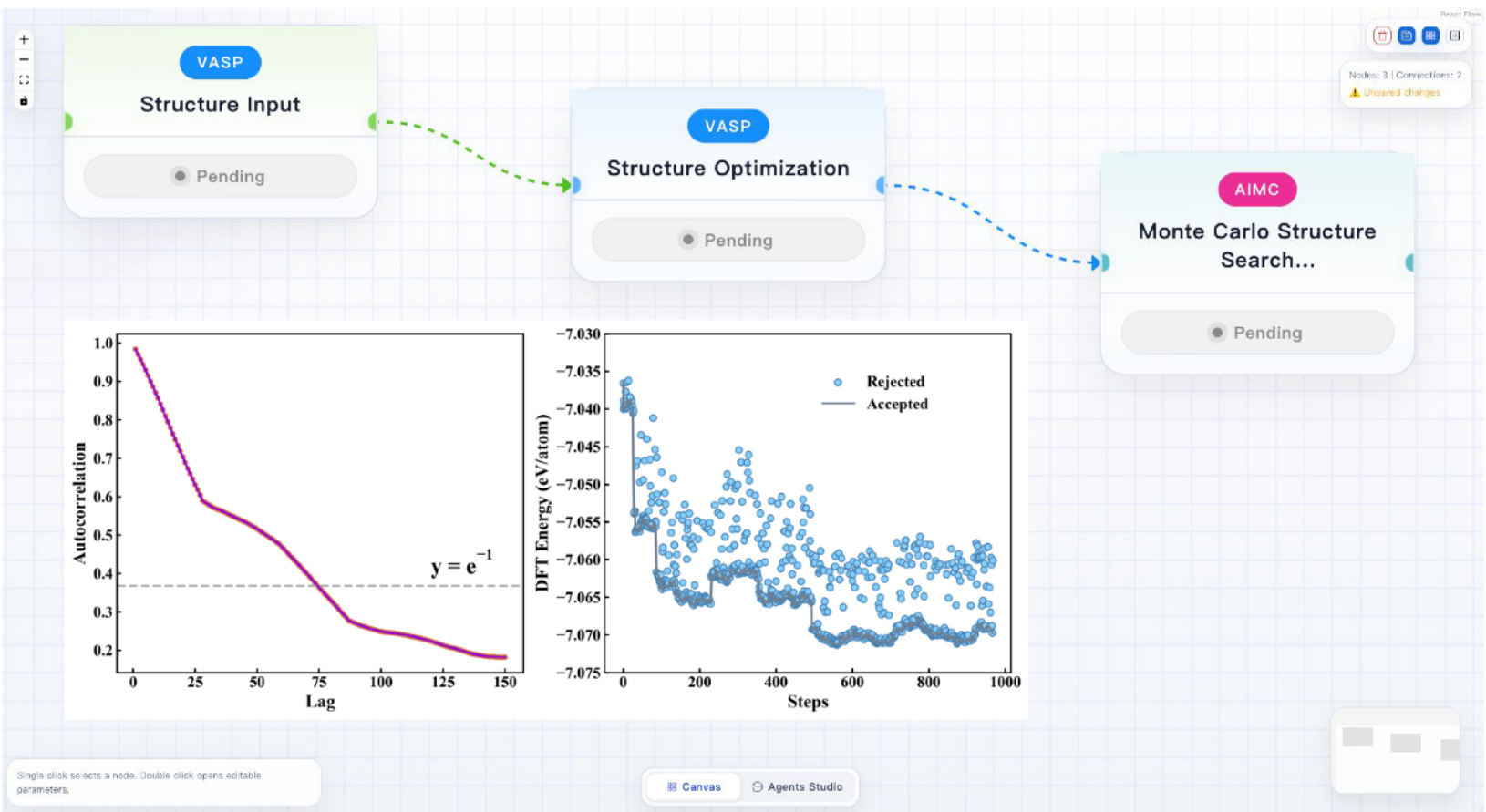

Fig. 20. AIMC task in ALKEMIE Agent for low-energy configuration sampling of multicomponent $(Re_{0.5}Nb_{0.5})(S_{0.5}Se_{0.5})_2$. The inset shows the autocorrelation function of the MC trajectory (left) and the energies of accepted and rejected atomic-exchange trials during configurational sampling (right).

### 4.7 Screening high-bulk-modulus materials using active learning

Here the active-learning module is assessed through a screening task for high-bulk-modulus materials based on the data from Materials Project[46,47]. As shown in Fig. 21, the module provides an integrated interface for defining the screening objective, selecting the surrogate model, setting the target threshold, specifying the batch size and exploration parameter, and enabling a random-selection branch for comparison. The source dataset contained 13,081 entries with available bulk-modulus labels. For the benchmark, 10% of the entries is randomly sampled as the initial labeled set. After removal of invalid or physically unreasonable values, 1,285 entries are retained for model training, while the remaining valid entries are treated as an unlabeled candidate pool. During selection, candidate-pool labels are hidden from the algorithm and are revealed only after each batch had been selected.

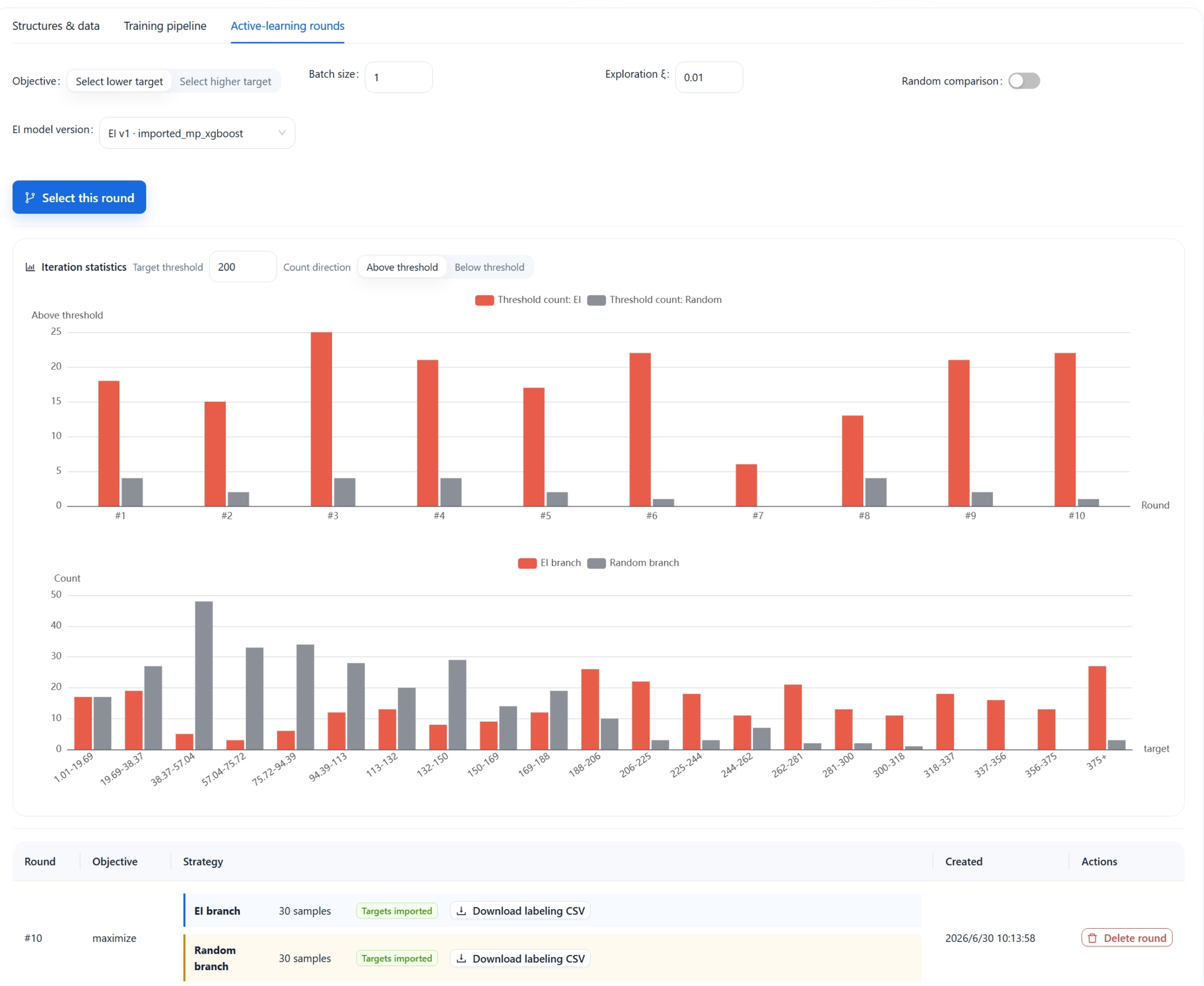

Fig. 21. User interface and diagnostic view of the active-learning module in the case of high bulk-modulus screening. Expected improvement is used to maximize bulk modulus with a 200 GPa target threshold, and a random-selection branch is included as a baseline. The diagnostic view compares round-level hit counts, selected-candidate distributions, and sample-labeling records for iterative model updating.

The bulk modulus above 200 GPa is defined as the target region, and expected improvement is used to maximize the target property. A random-selection branch is run in parallel under the same evaluation budget. Each strategy selected 30 structures per round for ten rounds, giving 300 evaluated candidates for each branch. The diagnostic results in Fig. 21 show that the EI branch selected more candidates above the 200 GPa threshold than the random branch and produced a distribution more concentrated in the high-bulk-modulus region. These results indicate that, under this benchmark setting, EI selection can direct the evaluation budget toward candidates with higher predicted bulk modulus. The interface also retains selected samples, revealed-label import status, labeling-file export, model information, and round records, thereby supporting traceable inspection and subsequent model updating.

## 5. Summary and Outlook

### 5.1 Summary

ALKEMIE Agent is introduced as a human-supervised agentic AI platform for computational materials design. The platform integrates task interaction, domain-knowledge retrieval, registered skills, context and task records, structure modeling, solver-facing execution, diagnostic assistance, and database-supported provenance within a common workflow environment. Through this integration, computational materials tasks can be represented as traceable task chains in which user intent, retrieved evidence, generated artifacts, execution states, parsed outputs, and provenance records remain connected.

The platform defines autonomy as bounded coordination rather than unrestricted scientific decision making. Task interpretation, route proposal, tool selection, input preparation, monitoring, parsing, and next-step suggestion are supported as observable platform actions. Calculation-defining choices, including structure construction, solver settings, convergence criteria, candidate acceptance, and final interpretation, remain subject to human review and approval.

To demonstrate the capability of ALKEMIE Agent, the presented examples cover materials recommendation for 2D magnetic systems, AI-assisted structure modeling and editing, VASP/Phonopy phonon calculation, GNEP task formulation and MLP-related workflow support, LAMMPS uniaxial tension simulation, AIMC atomic arrangement sampling, and active-learning screening for high-bulk-modulus materials. These cases indicate that computational materials workflows can be coordinated under the implemented task settings, with retained task records, execution traces, generated artifacts, parsed outputs, and provenance records supporting inspection, reproducibility, and reuse.

### 5.2 Outlook

Future updates of ALKEMIE Agent will emphasize workflow level self-iteration, cross-scale simulation and full-loop research capability. For the self-iteration, beyond the demonstrated active-learning-based iteration in material screening in which the surrogate model is updated automatically, result-driven workflow iteration will be introduced. This requires mechanisms by which parsed results are used to update candidates, inputs, parameters and calculation route, followed by traceable resubmission and result verification.

As for the cross-scale simulation, phase-field and finite-element methods are natural extensions of the execution-solvers in the future version of ALKEMIE Agent, for mesoscale and continuum materials modeling. The first challenge arises from data processing and data transfer across heterogeneous software packages. A second, more fundamental challenge concerns multi-fidelity data integration and uncertainty quantification. In particular, the propagation of errors from the quantum scale to the continuum level remains a critical issue. Ensuring that stochastic noise and approximation errors in lower-scale solvers do not induce non-physical artifacts in macroscopic predictions will require rigorous uncertainty-aware orchestration across the entire workflow.

A full-loop autonomous materials discovery, spanning hypothesis generation, simulation execution, analysis, and feedback-driven iteration, eventually integrating with autonomous experimental platforms, might be the ultimate goal for materials scientists. Such end-to-end workflows may become achievable in the near future for constrained materials systems under specific conditions; however, a key challenge remains in achieving generalizable functionality across diverse materials spaces and problem settings. Particularly, a fundamental class of challenges arises from inconsistencies between theoretical simulations and experimental observations. These include discrepancies in thermodynamic and kinetic conditions (e.g., idealized 0 K or equilibrium approximations in simulations versus finite-temperature, nonequilibrium experimental environments). Additional mismatches stem from incomplete or noisy experimental measurements, differences in sample quality (e.g., defects, disorder, impurities, and grain boundaries), and scale-dependent effects that are often absent in atomistic or electronic-structure models. Addressing these gaps will require robust theory-experiment coupling frameworks, improved uncertainty quantification, and data-driven calibration strategies that can reconcile multi-scale representations with real-world measurements.

Finally, systematic benchmarks are needed to evaluate route selection, input validity, parser accuracy, diagnostic quality, provenance completeness, numerical reproducibility, and stability across repeated runs for autonomous platforms such as ALKEMIE Agent. Establishing such benchmarks will likely require coordinated efforts from the broader research community.

## AUTHOR CONTRIBUTIONS

Linggang Zhu, Jian Zhou and Zhimei Sun proposed the project. Hongfu Huang designed the overall architecture of ALKEMIE Agent and developed the platform. Yuzhe Li and Tong Zhao contributed to the materials recommendation module. Ao Xu, Ning Yang, Hanyu Liu, Pengpeng Zhang, and Yichen Lu contributed to the computational software task modules. Bo Liu contributed to the note memory module. Changrui Wang contributed to the knowledge base. Shengxian Liu contributed to database construction. Kan Tang developed the active-learning module. Fengkai Liu contributed to the AI-assisted structure modeling and editing module. All the authors contributed to the optimization of the platform performance, analysis of the computed results, writing/revision of the manuscript.

## ACKNOWLEDGEMENTS

This work is financially supported by the Advanced Materials-National Science and Technology Major Project (2025ZD0618802).

## CONFLICT OF INTEREST STATEMENT

The authors declare no conflict of interest.

## DATA AVAILABILITY STATEMENT

A release version of ALKEMIE Agent is available at https://github.com/hfood02/alkemie-release. The current platform is available for

research use upon reasonable request to the corresponding author, subject to institutional policies, software-license restrictions, and computing-resource availability. Licensed solver executables, VASP pseudopotential files, user credentials, private project files, and site-specific HPC configuration files are not redistributed. A public web domain for broader researcher access is planned for a future release.

## References

1. Allison J. Integrated computational materials engineering: A perspective on progress and future steps. *JOM*. 2011;63(4):15-18. doi:10.1007/s11837-011-0053-y

2. Yang X, Wang Z, Zhao X, Song J, Zhang M, Liu H. MatCloud: A high-throughput computational infrastructure for integrated management of materials simulation, data and resources. *Computational Materials Science*. 2018;146:319-333. doi:10.1016/j.commatsci.2018.01.039

3. Van Der Spoel D, Lindahl E, Hess B, Groenhof G, Mark AE, Berendsen HJC. GROMACS: Fast, flexible, and free. *J Comput Chem*. 2005;26(16):1701-1718. doi:10.1002/jcc.20291

4. Behler J. Atom-centered symmetry functions for constructing high-dimensional neural network potentials. *The Journal of Chemical Physics*. 2011;134(7):074106. doi:10.1063/1.3553717

5. Hafner J. Ab-initio simulations of materials using VASP: Density-functional theory and beyond. *Journal of Computational Chemistry*. 2008;29(13):2044-2078. doi:10.1002/jcc.21057

6. Kresse G. *Ab initio* molecular-dynamics simulation of the liquid-metal–amorphous-semiconductor transition in germanium. *Phys Rev B*. 1994;49(20):14251-14269. doi:10.1103/PhysRevB.49.14251

7. Ziesche P, Kurth S, Perdew JP. Density functionals from LDA to GGA. *Computational Materials Science*. 1998;11(2):122-127. doi:10.1016/S0927-0256(97)00206-1

8. Kühne TD, Iannuzzi M, Del Ben M, et al. CP2K: An electronic structure and molecular dynamics software package - Quickstep: Efficient and accurate electronic structure calculations. *J Chem Phys*. 2020;152(19). doi:10.1063/5.0007045

9. Huang H, Peng J, Li K, Zhou J, Sun Z. Efficient GPU-accelerated training of a neuroevolution potential with analytical gradients. *Computer Physics Communications*. 2026;320:109994. doi:10.1016/j.cpc.2025.109994

10. Steinbach I. Phase-field models in materials science. *Modelling Simul Mater Sci Eng*. 2009;17(7):073001. doi:10.1088/0965-0393/17/7/073001

11. Roters F, Eisenlohr P, Hantcherli L, Tjahjanto DD, Bieler TR, Raabe D. Overview of constitutive laws, kinematics, homogenization and multiscale methods in crystal plasticity finite-element modeling: Theory, experiments, applications. *Acta Materialia*. 2010;58(4):1152-1211. doi:10.1016/j.actamat.2009.10.058

12. Ong SP, Richards WD, Jain A, et al. Python Materials Genomics (pymatgen): A robust, open-source python library for materials analysis. *Computational Materials Science*. 2013;68:314-319. doi:10.1016/j.commatsci.2012.10.028

13. Wang G, Peng L, Li K, et al. ALKEMIE: An intelligent computational platform for accelerating materials discovery and design. *Computational Materials Science*. 2021;186:110064. doi:10.1016/j.commatsci.2020.110064

14. Huber SP, Zoupanos S, Uhrin M, et al. AiiDA 1.0, a scalable computational infrastructure for automated reproducible workflows and data provenance. *Sci Data*. 2020;7(1):300. doi:10.1038/s41597-020-00638-4

15. Calderon CE, Plata JJ, Toher C, et al. The AFLOW standard for high-throughput materials science calculations. *Computational Materials Science*. 2015;108:233-238. doi:10.1016/j.commatsci.2015.07.019

16. Kirklin S, Saal JE, Meredig B, et al. The Open Quantum Materials Database (OQMD): assessing the accuracy of DFT formation energies. *npj Comput Mater*. 2015;1(1):15010. doi:10.1038/npjcompumats.2015.10

17. Mathew K, Montoya JH, Faghaninia A, et al. Atomate: A high-level interface to generate, execute, and analyze computational materials science workflows. *Computational Materials Science*. 2017;139:140-152. doi:10.1016/j.commatsci.2017.07.030

18. Jain A, Ong SP, Chen W, et al. FireWorks: a dynamic workflow system designed for high-throughput applications. *Concurrency and Computation*. 2015;27(17):5037-5059. doi:10.1002/cpe.3505

19. Miret S, Krishnan NMA. Enabling large language models for real-world materials discovery. *Nat Mach Intell*. 2025;7(7):991-998. doi:10.1038/s42256-025-01058-y

20. Pyzer-Knapp EO, Manica M, Staar P, et al. Foundation models for materials discovery – current state and future directions. *npj Comput Mater*. 2025;11(1):61. doi:10.1038/s41524-025-01538-0

21. Jiang X, Wang W, Tian S, Wang H, Lookman T, Su Y. Applications of natural language processing and large language models in materials discovery. *npj Comput Mater*. 2025;11(1):79. doi:10.1038/s41524-025-01554-0

22. Ghafarollahi A, Buehler MJ. AtomAgents: Alloy design and discovery through physics-aware multi-modal multi-agent artificial intelligence.

23. Zhang B, Li X, Xu H, Jin Z, Wu Q, Li C. TopoMAS: Large Language Model Driven Topological Materials Multi-Agent System. *Materials Genome Engineering Advances*. 2026;4(1):e70045. doi:10.1002/mgea.70045

24. Chaudhari A, Ock J, Barati Farimani A. Modular large language model agents for multi-task computational materials science. *Commun Mater*. 2026;7(1):131. doi:10.1038/s43246-025-00994-x

25. Liu J, Zhu T, Ye C, Fang Z, Weng H, Wu Q. VASPilot: MCP-Facilitated Multi-Agent Intelligence for Autonomous VASP Simulations. *arXiv*. Preprint posted online 2025. doi:10.48550/ARXIV.2508.07035

26. Zhang H, Song Y, Hou Z, Miret S, Liu B. HoneyComb: A Flexible LLM-Based Agent System for Materials Science. In: *Findings of the Association for Computational Linguistics: EMNLP 2024*. Association for Computational Linguistics; 2024:3369-3382. doi:10.18653/v1/2024.findings-emnlp.192

27. Wang X, Zeng Q, Xu DH, Zhang L, Jiang G, Yang M. Accelerating materials discovery via AI-Agent integration of large language models and simulation tools. *jmi*. 2026;6(1):N/A-N/A. doi:10.20517/jmi.2025.69

28. Ding Q, Miret S, Liu B, Courtois I. MATEXPERT: DECOMPOSING MATERIALS DISCOVERY BY MIMICKING HUMAN EXPERTS. Published online 2025.

29. Wang X, Li C, Zhang B, et al. S1-MatAgent: A planner driven multi-agent system for material discovery.

30. Li C, Ran N, Liu J. Agentic material science. *J Mater Inf*. 2026;6(1). doi:10.20517/jmi.2025.87

31. Shi Z, Xin C, Huo T, et al. A fine-tuned large language model based molecular dynamics agent for code generation to obtain material thermodynamic parameters. *Sci Rep*. 2025;15(1):10295. doi:10.1038/s41598-025-92337-6

32. Yao Z, Zhang B, Shu J, et al. MatMind: A Structure-Activity Knowledge-Driven Generative Foundation Model for Materials Science. *arXiv*. Preprint posted online June 5, 2026:arXiv:2606.07712. doi:10.48550/arXiv.2606.07712

33. Chiang Y, Hsieh E, Chou CH, Riebesell J. LLaMP: Large Language Model Made Powerful for High-fidelity Materials Knowledge Retrieval and Distillation. *arXiv*. Preprint posted online October 9, 2024:arXiv:2401.17244. doi:10.48550/arXiv.2401.17244

34. Ghafarollahi A, Buehler MJ. Automating alloy design and discovery with physics-aware multimodal multiagent AI. *Proc Natl Acad Sci USA*. 2025;122(4):e2414074122. doi:10.1073/pnas.2414074122

35. Ward L, Dunn A, Faghaninia A, et al. Matminer: An open source toolkit for materials data mining. *Computational Materials Science*. 2018;152:60-69. doi:10.1016/j.commatsci.2018.05.018

36. Chen T, Guestrin C. XGBoost: A Scalable Tree Boosting System. In: *Proceedings of the 22nd ACM SIGKDD International Conference on Knowledge Discovery and Data Mining*. ACM; 2016:785-794. doi:10.1145/2939672.2939785

37. Jones DR, Schonlau M. Efficient Global Optimization of Expensive Black-Box Functions.

38. Lookman T, Balachandran PV, Xue D, Yuan R. Active learning in materials science with emphasis on adaptive sampling using uncertainties for targeted design. *npj Comput Mater*. 2019;5(1):21. doi:10.1038/s41524-019-0153-8

39. Balachandran PV, Xue D, Theiler J, Hogden J, Lookman T. Adaptive Strategies for Materials Design using Uncertainties. *Sci Rep*. 2016;6(1):19660. doi:10.1038/srep19660

40. Togo A, Tanaka I. First principles phonon calculations in materials science. *Scripta Materialia*. 2015;108:1-5. doi:10.1016/j.scriptamat.2015.07.021

41. van de Walle A, Tiwary P, de Jong M, et al. Efficient stochastic generation of special quasirandom structures. *Calphad*. 2013;42:13-18. doi:10.1016/j.calphad.2013.06.006

42. Yang N, Zhu L, Liu H, Zhou J, Sun Z. Enhancing the mechanical properties of TiZr-based multi-principal element alloys via leveraging multiple short-range orders: An atomic-scale study. *Journal of Materials Science & Technology*. 2025;227:133-141. doi:10.1016/j.jmst.2024.11.063

43. Zhang T, Zhu L, Liu H, Zhou J, Sun Z. Ordering phenomena in ternary transition-metal dichalcogenides: Critical role of lattice symmetry and vdW interaction. *Materials Genome Engineering Advances*. 2023;1(2):e7. doi:10.1002/mgea.7

44. Liu H, Zhu L, Zhou J, Sun Z. Deciphering Short-Range Order in 2D Transition Metal Dichalcogenides: From Origin to Multi-Scale Property Modulation. *Advanced Science*. Published online April 17, 2026:e24378. doi:10.1002/advs.202524378

45. Liu H, Zhu L, Zhou J, Sun Z. Competing sublattice short-range orders and gap state engineering in multicomponent transition-metal dichalcogenide. *npj Comput Mater*. 2025;12(1):30. doi:10.1038/s41524-025-01899-6

46. Jain A, Ong SP, Hautier G, et al. Commentary: The Materials Project: A materials genome approach to accelerating materials innovation. *APL Materials*. 2013;1(1):011002. doi:10.1063/1.4812323

47. De Jong M, Chen W, Angsten T, et al. Charting the complete elastic properties of inorganic crystalline compounds. *Sci Data*. 2015;2(1):150009. doi:10.1038/sdata.2015.9